\documentclass[aps,notitlepage,longbibliography]{revtex4-1}

\usepackage{CJK}
\usepackage{tabularx}
\usepackage{graphicx}
\usepackage{bm}

\usepackage[
margin=1.2in,
]{geometry}
\usepackage{array}
\usepackage{makecell}
\usepackage{natbib}
\usepackage{float}
\usepackage{todo}
\usepackage{enumitem}
\usepackage{multirow}
\usepackage{tabularx}
\usepackage{gensymb}
\usepackage{footnote}
\usepackage{amsmath}

\def\bdim{b^*}
\def\hdim{h^*_{\infty}}

\usepackage{tikz}
\usetikzlibrary{arrows.meta}

\usepackage{hyperref}
\hypersetup{
    colorlinks,
    citecolor=black,
    filecolor=black,
    linkcolor=black,
    urlcolor=black}

\usepackage{xcolor}
\newcommand{\red}[1]{{\color{black}#1}}

\newcommand{\pdv}[2]{\frac{\partial #1}{\partial #2}}
\newcommand{\pdvstar}[3][]{\frac{\partial^{#1} #2}{\partial #3 ^{*#1}}}

\begin{document}

\begin{CJK*}{GB}{} 

\title[Undercompressive shocks in tears of wine]{A theory for undercompressive shocks in tears of wine}
\author{Yonatan Dukler$^{*}$}
\author{Hangjie Ji$^{*}$}
\author{Claudia Falcon$^{*}$}
\author{Andrea L. Bertozzi$^{*\dagger}$}
\affiliation{$^*$Department of Mathematics}

\affiliation{$\dagger$ Department of Mechanical and Aerospace Engineering Department, University of California, Los Angeles, Los Angeles, California 90095, USA}
\email{ [ ydukler, bertozzi, hangjie, cfalcon ]@math.ucla.edu }


\begin{abstract}
We revisit the tears of wine problem for thin films in water-ethanol mixtures and present a model for the climbing dynamics. 
The new formulation includes a Marangoni stress balanced by both the normal and tangential components of gravity as well as surface tension which lead to distinctly different behavior.  The prior literature did not address the wine tears but rather the behavior of the film at earlier stages and the behavior of the meniscus.
In the lubrication limit we obtain an equation that is already well-known for rising films in the presence of thermal gradients. Such models can exhibit nonclassical shocks that are undercompressive.  
We present basic theory that allows one to identify the signature of an undercompressive wave.
We observe both compressive and undercompressive waves in new experiments and we argue that, in the case of a preswirled glass, the famous ``wine tears'' emerge from a reverse undercompressive shock originating at the meniscus.

\end{abstract}

\maketitle

\end{CJK*}

\section{\label{sec:level1}Introduction}

The tears of wine problem is a curious \red{phenomenon} that has been observed in wine glasses for centuries. In the right setting, one can observe a thin layer of water-ethanol mixture that travels up inclined surfaces against gravity and proceeds to fall down in the form of tears. This is a result of a solutal Marangoni stress counterbalanced with the force of gravity.
The Marangoni stress stems from a surface tension gradient caused by alcohol evaporation and the resulting difference in alcohol concentration.
Specifically, when a water-ethanol mixture is placed in a container with inclined walls, a thin meniscus forms.
The alcohol in the meniscus region becomes more depleted than that of the bulk due to the predominant role of evaporation in the meniscus.
This leads to a solutal surface tension gradient that pulls liquid out of the meniscus and up the side of the glass.

The phenomenon of wine tears was first analyzed qualitatively by Thomson \cite{thomson1855xlii} who attributed it to the Marangoni stress.
In 1992 the first careful experiments were conducted by Fournier and Cazabat to understand the phenomenon \cite{fournier1992tears}. Further studies focused on the various instabilities that form \cite{vuilleumier1995tears, hosoi2001evaporative, fanton1998spreading}. In particular the work of Vuilleumier \textit{et al.} \cite{vuilleumier1995tears} focuses on the stationary state when the film reaches its terminal height and star instabilities form in addition to the tears. In the paper by Fanton and Cazabat \cite{fanton1998spreading} studying spreading instabilities, the authors continue describing the star instabilities that form in two component mixtures. In 2001, Hosoi and Bush \cite{hosoi2001evaporative} further investigated two distinct instabilities in the climbing film using a lubrication model that includes gravity, capillarity and Marangoni stresses.
The work of Venerus and Simavilla \cite{venerus2015tears} identifies a previously overlooked temperature gradient due to evaporation, that also contributes to the Marangoni stress. More recently, Nikolov \textit{et al.}\cite{nikolov2018tears} also applied the Plateau-Rayleigh-Taylor theory to study the ridge instability that triggers the formation of wine tears.  However the dynamic formation of the ridge is still not well-understood.
\par

\red{ All of the prior works on tears of wine neglected to consider the tangential component of
gravity along with the other physics.  The tangential Marangoni stress,
tangential component of
gravity, and the bulk surface tension lead to a dynamic model that is
known to produce unusual behavior sometimes characterized by nonclassical
shocks.  This has been well-studied in thermally driven films \cite{munch2006interaction,munch1999rarefaction,munch2003pinch,bertozzi1999undercompressive,bertozzi1998contact} but never in
tears of wine.  In
models studied in the literature \cite{hosoi2001evaporative, vuilleumier1995tears,venerus2015tears}, one expects a moving front
with advancing fingers, which
is inconsistent with the draining tears observed in experiments. This
suggests that a more intricate
mechanism is taking place, motivating further studies.}

Via an enhanced model, we illustrate the existence of nonclassical undercompressive shocks for the first time in the context of tears of wine. 
This model better characterizes the dynamics of climbing films which sheds light on the experimental work in the literature. 
Relevant to our analysis are the works studying the structure and shock formation in thermally-driven thin films where nonclassical shocks have been observed \cite{munch2006interaction,munch1999rarefaction,bertozzi1999undercompressive,bertozzi1998contact}. 
In this paper, we investigate different shock morphologies that can spontaneously occur in climbing films of wine, depending on the experimental settings and alcohol concentration. 
For instance, we expect undercompressive shocks in the experiments of \cite{vuilleumier1995tears}. \par
\red{More importantly, we take a closer look at the common wine glass setting, something not well-studied in prior works.} This corresponds to using a radially symmetric glass, and incorporating swirling as done in common handling of wine. We find that the geometry and swirling of the glass affect the formation of tears, which differs from the better-studied studied spontaneous climb. Mathematically, the new setting translates to extending the model to incorporate additional geometries, and adding a pre-swirling draining fluid layer. Specifically, our analysis shows that the draining fluid can give rise to reverse undercompressive shocks \cite{munch2003pinch} that help explain the formation of tears from a climbing reverse front, which we find to be quite reproducible, experimentally, with steeper beverage glasses and higher alcohol concentrations.

The rest of the paper is structured as follows: 
In section \ref{sec:level2} we lay out the theory, deriving the non-dimensional PDE model for the climbing thin film.
The shape of the meniscus and front dynamics, in addition to the relevant works on the mathematical theory of undercompressive shocks in thin films are introduced in section \ref{sec:shocktheory}.
In section \ref{sec:survey} we review the experimental work in the literature, and present numerical simulations of our new model using corresponding experimental parameters. The effects of glass geometries on the film dynamics are investigated in section \ref{sec:curvatureEffects}.
The appearance of an unusual reverse undercompressive shock wave triggered by a draining film is discussed in section \ref{sec:RUC}.
Lastly we discuss our findings on different shocks and hypothesize their relation to the formation of tears in section \ref{sec:conclusion}.

\section{Hydrodynamic model}
\label{sec:level2}
\begin{figure}
\centering
\includegraphics[width=4.5cm, height = 5cm]{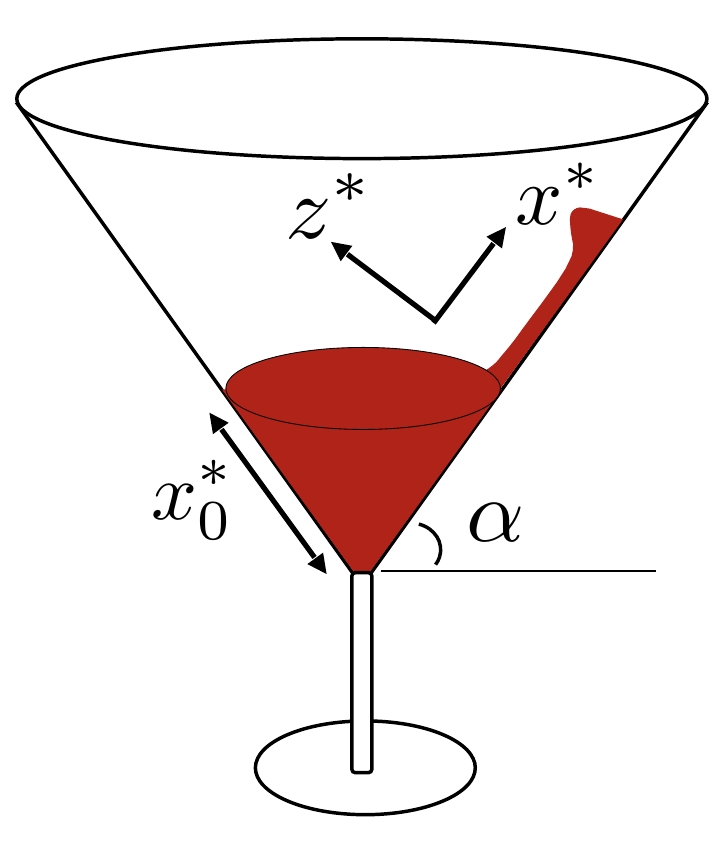}\qquad
\includegraphics[width = 0.6 \textwidth]{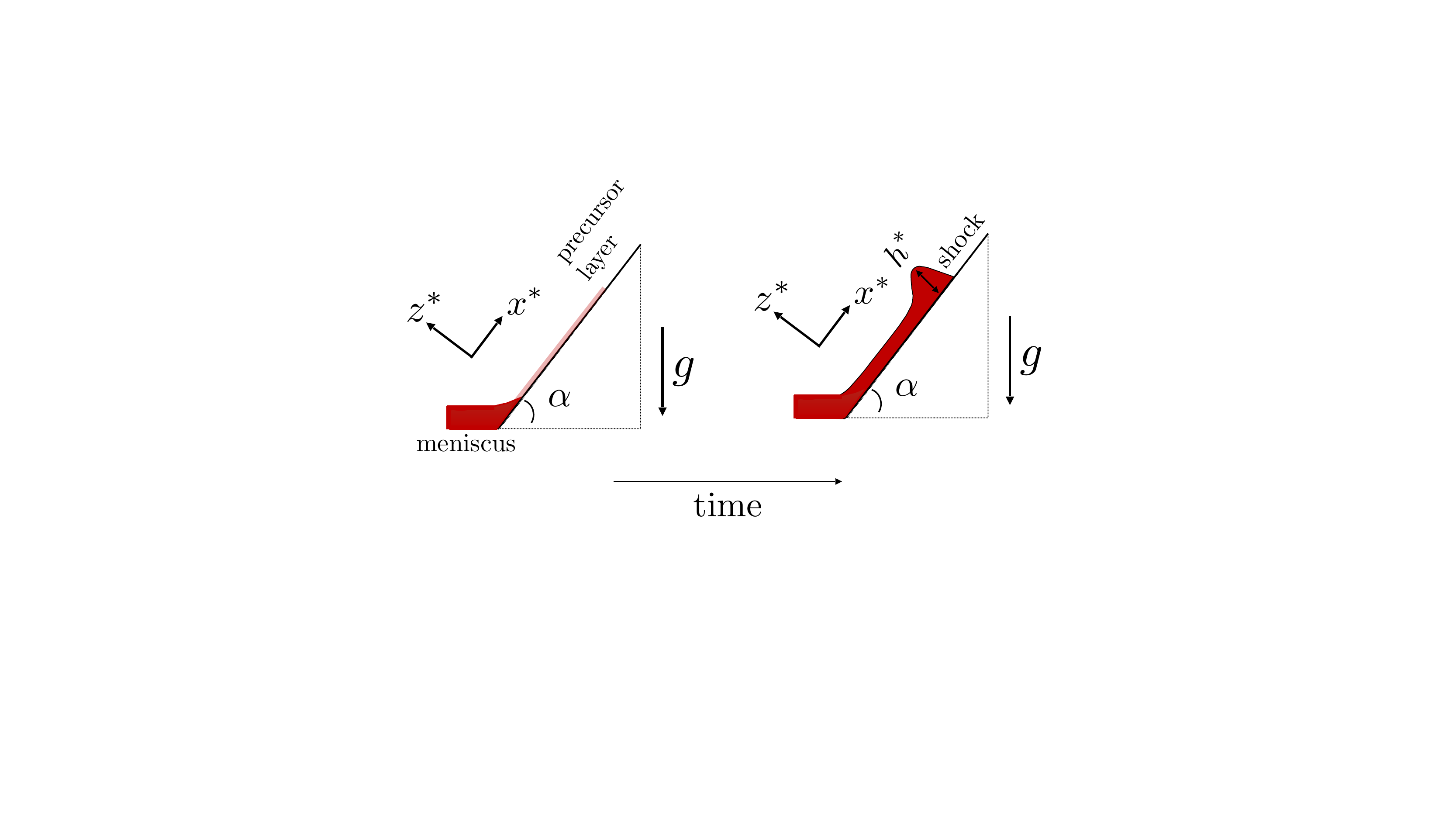}
\caption{(left) Schematic illustration of a conical-shaped cocktail glass of inclination angle $\alpha$, and (right) the corresponding one-dimensional thin wine film travelling up inclined flat glass surface. The film height $h^*$ is exaggerated for clarity of the illustration. 
}
\label{fig:towlabel}
\end{figure}
We derive our model building on the foundational model presented in the work of Fournier and Cazabat \cite{fournier1992tears}. \red{Based on conservation of mass of the liquid the authors derive the following equation for the thin film flux}
\red{\begin{equation}
    \frac{\partial h^*}{\partial t^*} + \frac{\partial Q^*}{\partial x^*}  = 0, \qquad Q^* = h^* v^*,
\end{equation}}
where \red{$h^*(x^*,t^*)$} is the \red{dimensional} film thickness, $v^*$ is the \red{average velocity across the film}, \red{and} $Q^*$ is the flux.
Then the velocity is written in terms of the \red{surface tension $\gamma$} and the dynamic viscosity $\mu$, \red{representing convection of the film due to the surface tension gradient}
\begin{equation}
 v^* = \frac{h^*}{2\mu} \pdv{\gamma}{x^*},
 \label{eq:simple_model}
\end{equation}
We incorporate the tangential and normal components of gravity and the surface tension to \red{the model in equation \eqref{eq:simple_model} and obtain}
\begin{equation}
v^* = \frac{h^*}{2\mu}\pdv{\gamma}{x^*}- \frac{h^{*2}}{3}\bigg(\frac{\rho g \sin{\alpha}}{\mu} +\frac{\rho g\cos{\alpha}}{\mu} \pdv{h^*}{x^*} - \frac{\gamma}{\mu}\pdvstar[3]{h^*}{x}\bigg),
\label{eq:v}
\end{equation} 
where $g$ is gravity, $\rho$ is density, $\alpha$ is the inclination angle of the surface and $\gamma$ is the surface tension of the film. 
\red{The formula \eqref{eq:v} for $v^*$ comes from the lubrication theory \cite{oron1997long,fanton1996thickness,carles1993thickness,bertozzi1999undercompressive},  which is a long wavelength approximation of the classic Navier-Stokes equations in the low Reynolds number limit. In addition to the first term with surface tension gradient $\partial \gamma/\partial x^*$ from the formula \eqref{eq:simple_model}, the second term in \eqref{eq:v} represents the convection of the film due to the component of gravity tangential to the surface, the $\partial h^*/\partial x^*$ term represents the diffusion of the film caused by the normal component of gravity, and the last term with $\partial^3 h^*/\partial x^{*3}$ comes from the surface tension.}
Using the enhanced model the flux is then reformulated as
\begin{equation}
Q^* = \frac{h^{*2}}{2\mu}\pdv{\gamma}{x^*}- \frac{h^{*3}}{3}\bigg(\frac{\rho g \sin{\alpha}}{\mu} +\frac{\rho g \cos{\alpha}}{\mu} \pdvstar{h^*}{x} - \frac{\gamma}{\mu}\pdvstar[3]{h^*}{x}\bigg).
\label{eq:enhanced_flux}
\end{equation} 

For simplicity we assume a constant surface tension gradient $\tau$ following prior works \cite{fournier1992tears, vuilleumier1995tears, hosoi2001evaporative}.
Our model then reduces to

\begin{equation}
\frac{\partial h^*}{\partial t^*}+ \frac{\tau}{2\mu}\frac{\partial}{\partial x^*}\left(h^{*2}\right) -
\frac{\partial}{\partial x^*}\bigg(\frac{h^{*3}}{3}\frac{g\rho\sin{\alpha}}{\mu} \bigg)  = 
\frac{\partial}{\partial x^*}\bigg[ \frac{h^{*3}}{3}\bigg(\frac{g\rho \cos{\alpha}}{\mu} \pdv{h^*}{x^*} - \frac{\gamma}{\mu}\pdvstar[3]{h^*}{x}\bigg) \bigg].
\label{eq:dim}
\end{equation}

By balancing the Marangoni stress term and the tangential component of the gravity,
we then non-dimensionalize the PDE as in the work of M{\"u}nch and Evans  \cite{munch2006interaction} 
using \red{
\begin{equation}
    h^* = Hh,\qquad  x^* = Xx,\qquad t^* = Tt,
    \nonumber
\end{equation}
where
}
\begin{equation}
  H =\displaystyle{\frac{3\tau}{2g\rho \sin{\alpha}}},\qquad
X = \sqrt[3]{\frac{3\gamma \tau}{2(\rho g \sin{\alpha})^2}},\qquad
 T = \displaystyle{2 \mu \sqrt[3]{\frac{4\gamma \rho g \sin{\alpha}}{9\tau^5}}},
 \label{scaling}
\end{equation}
which gives the non-dimensional equation
\begin{equation}
  h_t + \left[f(h)\right]_{x} = -(h^3h_{xxx})_x + D(h^3h_x)_x.
\label{eqn:nondim}
\end{equation}
\red{Here we denote $h,x$ for the dimensionless film height and length.}
The constant $D$ is defined as
\begin{equation}
  D = \sqrt[3]{\frac{9 \tau^2\cos^3{\alpha}}{4\gamma \rho g \sin^{4}{\alpha}}},
  \label{eq:D}
\end{equation}
and the non-convex flux function $f(h)$ takes the form
\begin{equation}
 f(h) = h^2-h^3,
 \label{eq:flux}  
\end{equation}
where the quadratic and cubic terms come from the Marangoni stress and the tangential component of the gravity, respectively.
This equation has been studied in thermally driven films and can exhibit nonclassical shocks in some regimes \red{\cite{munch2006interaction}}. It is worth mentioning that our formulation allows for a general surface tension gradient which may be the result of solutal and thermal surface tension gradients as pointed out by Venerus and Simavilla \cite{venerus2015tears}. 
Using experimental data provided in the literature, we note that the non-dimensional constant $D = O(1)$, which indicates that the added term that corresponds to the gravity in the normal direction is necessary to capture the full dynamics. 

Extensive studies have shown that the interaction between the non-convex flux function and the higher-order smoothing term in \eqref{eqn:nondim} can lead to interesting shock wave structures \cite{bertozzi1999undercompressive}. For the rest of the paper, we focus on the analysis of different shock formation mechanisms in two separate cases: the spontaneous climbing of wine films in a static glass, and the climbing film with the presence of a draining film after a glass swirling.\par

\begin{figure}
    \centering
        
  \begin{tabular}{c c c}
        {
        \hbox{ 
         \resizebox{4.2cm}{3.6cm}{
    \begin{tikzpicture}[scale=0.7]
    \node[draw] at (3.5,4.5) {C};
    \draw[line width =0.4mm,] (2,0) -- (2,5);
    \draw[line width =0.3mm,-{Latex[length=3mm, width=1.5mm]}] (1,0) -- (2,2);
    \draw[line width =0.3mm,-{Latex[length=3mm, width=1.5mm]}] (3,0) -- (2,2);
    \draw[line width =0.3mm,-{Latex[length=3mm, width=1.5mm]}] (1.5,0) -- (2,1);
    \draw[line width =0.3mm,-{Latex[length=3mm, width=1.5mm]}] (2.5,0) -- (2,1);
    \draw[line width =0.3mm,-{Latex[length=3mm, width=1.5mm]}] (3.5,0) -- (2,3);
    \draw[line width =0.3mm,-{Latex[length=3mm, width=1.5mm]}] (0.5,0) -- (2,3);
    \draw[line width =0.3mm,-{Latex[length=3mm, width=1.5mm]}] (4,0) -- (2,4);
    \draw[line width =0.3mm,-{Latex[length=3mm, width=1.5mm]}] (0,0) -- (2,4);

    \draw[loosely dotted] (0,0) grid (4,5);
    \path[use as bounding box] (0,0) rectangle (4,4);
    \draw[->] (-0.2,0) -- (4.25,0) node[right] {\small
 $x$};
    \draw[->] (0,-0.25) -- (0,5.25) node[above] { \small $t$};
    \end{tikzpicture}
    }
    }
    }
    
     &
    {
     \hspace{0.2in}
     \resizebox{4.2cm}{3.6cm}{
          \begin{tikzpicture}[scale=0.7]
    \node[draw] at (3.2,4.5) {UC};
    \draw[line width =0.4mm,] (2,0) -- (2,5);
    \draw[line width =0.3mm,-{Latex[length=3mm, width=1.5mm]}] (2,2)--(0,3);
    \draw[line width =0.3mm,-{Latex[length=3mm, width=1.5mm]}] (3,0) -- (2,2);
    \draw[line width =0.3mm,-{Latex[length=3mm, width=1.5mm]}]  (2,1)--(0,2);
    \draw[line width =0.3mm,-{Latex[length=3mm, width=1.5mm]}] (2.5,0) -- (2,1);
    \draw[line width =0.3mm,-{Latex[length=3mm, width=1.5mm]}] (3.5,0) -- (2,3);
    \draw[line width =0.3mm,-{Latex[length=3mm, width=1.5mm]}]  (2,3)--(0,4);
    \draw[line width =0.3mm,-{Latex[length=3mm, width=1.5mm]}] (4,0) -- (2,4);
     \draw[line width =0.3mm,-{Latex[length=3mm, width=1.5mm]}] (2,4) -- (0,5);
          \draw[line width =0.3mm,-{Latex[length=3mm, width=1.5mm]}] (2,0) -- (0,1);
          
    \draw[loosely dotted] (0,0) grid (4,5);
    \path[use as bounding box] (0,0) rectangle (4,4);
    \draw[->] (-0.2,0) -- (4.25,0) node[right] {\small $x$};
    \draw[->] (0,-0.25) -- (0,5.25) node[above] {\small $t$};

    \end{tikzpicture}
    }} 
    & {
        \hspace{-0.1in}
        \includegraphics[width=6cm, height = 4.2cm]{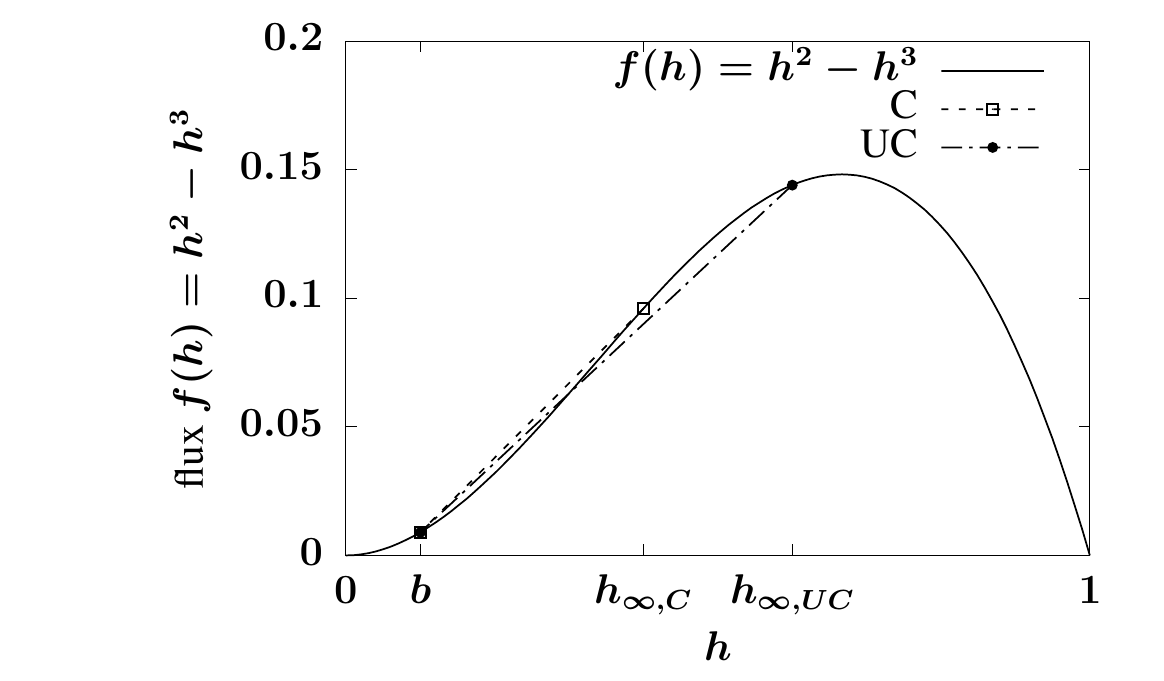}
     } 
\end{tabular}  
\caption{Schematics for the characteristics of (left) a compressive shock \red{that satisfies \eqref{eq:lax}} and (middle) an undercompressive shock \red{that satisfies \eqref{eq:notlax}} shown in a moving reference frame in which the shock is stationary. (right) Flux function $f(h)$ in \eqref{eq:flux} with a compressive connection from $h_{\infty, C} = 0.4$ to $b = 0.1$ and an undercompressive connection from $h_{\infty, UC} = 0.6$ to $b = 0.1$ \red{where $h_{\infty}$ and $b$ are the left and right boundary conditions of equation \eqref{eqn:nondim}.}
}
 \label{fig:characteristics}
\end{figure}
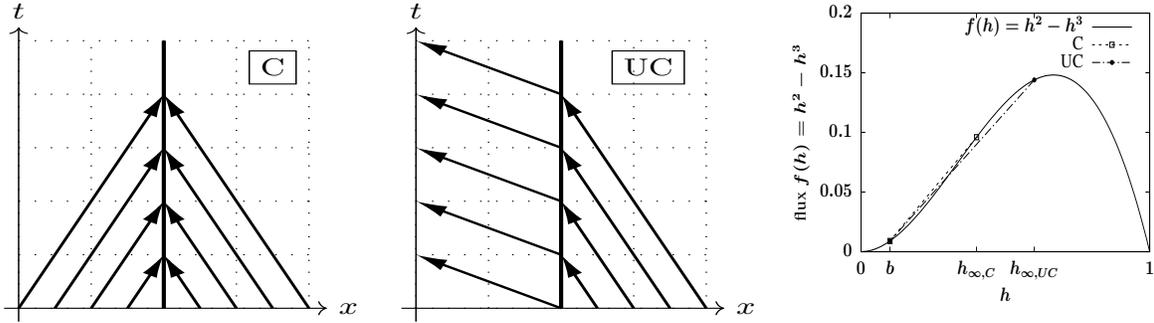

    We briefly review the nonlinear dynamics of classical compressive and nonclassical undercompressive shocks. For simplicity, we focus on a single shock that arises from equation \eqref{eqn:nondim} coupled to the far-field boundary conditions
    \begin{equation}
        h \to h_{\infty} \text{ as } x \to -\infty, \qquad h \to b \text{ as } x \to +\infty,
        \label{flatBC}
    \end{equation}
\red{$h_{\infty}$ and $b$ are the left and right boundary conditions of the solution.}
When analyzing equation \eqref{eqn:nondim} we consider the travelling wave solutions of the form
$
h(x,t) = \bm\hat{h}(x-st)
$,
where $s$ is the speed of the wave. Adjusting to the reference frame of the shock, the flux can be written as 
$
\bm\hat{f}(\bm\hat{h}) =  \bm\hat{h}^2 - \bm\hat{h}^3  -s \bm\hat{h},
$
which controls  $\bm\hat{h}$ via the ODE
\[
\left[\bm\hat{f}(\bm\hat{h})\right]_{x}  -(\bm\hat{h}^3\bm\hat{h}_{xxx})_x + D(\bm\hat{h}^3\bm\hat{h}_x)_x = 0.
\]
Integrating this equation using the left and right boundary conditions \eqref{flatBC} gives the standard Rankine-Hugoniot jump condition for the speed of the shock $
s  = \left(f(h_{\infty})-f(b)\right)/\left(h_{\infty} - b\right).
$
On the other hand, for large time and space scales one may drop the higher order diffusive terms in equation \eqref{eqn:nondim} which leads to the quasi-linear hyperbolic equation
\begin{equation}
    h_t + [f(h)]_x = 0.
    \label{scaler_hyperbolic}
\end{equation}
This reduced equation yields the speed of the characteristics $f'(h) = 2h - 3h^2$. With $\eta = x/t$, PDE \eqref{scaler_hyperbolic} also admits solutions that consists of an expanding rarefaction wave,
\begin{equation}
    h(x,t) = H(\eta), \qquad \mbox{where}\quad H(\eta) = (f')^{-1}(\eta) =  \textstyle{\frac{1}{3}}-\textstyle{\frac{1}{3}}\sqrt{1-3\eta}.
    \label{rarefaction}
\end{equation}

The Lax entropy condition for compressive shocks is given as 
\begin{equation}
f'(b) < s < f'(h_{\infty}),
\label{eq:lax}
\end{equation}
or in the moving reference of speed $s$, $    \bm\hat{f}'(b) < 0 < \bm\hat{f}'(h_{\infty}).
$
 A characteristic diagram for a compressive shock in the moving reference is illustrated in FIG.~\ref{fig:characteristics} (left) with characteristics entering from both sides of the shock. This type of shock can also be identified as a chord connecting the left and right states of the shock in a flux diagram. One such example is shown in FIG.~\ref{fig:characteristics} (right) where a chord connects the left state $h_{\infty, C} = 0.4$ and the right state $b = 0.1$ of a compressive shock.

Interestingly, for undercompressive shocks the Lax condition \eqref{eq:lax} is violated with 
\begin{align}
f'(b) < f'(h_{\infty}) < s,
\label{eq:notlax}
\end{align}
or in the moving reference of speed $s$, $\bm\hat{f}'(b) < \bm\hat{f}'(h_{\infty}) < 0 .$ 
This is visualized in FIG.~\ref{fig:characteristics} (middle) where the characteristics travel through the shock,
with the undercompressive connection from $h_{\infty, UC} = 0.6$ to $b = 0.1$ plotted in FIG.~\ref{fig:characteristics} (right).

Information propagating through the undercompressive shocks correspond to stability to traverse perturbations \cite{bowen2005nonlinear}. This stability is a mark of undercompressive shocks that does not occur in classical compressive shocks and will be used in distinguishing compressive and undercompressive shocks in the fluid experiments.

Stability of the shock may be analyzed by considering the properties of the perturbed solution,
$
    h(x,t) = \red{\tilde{h}_0(x,t)} + \epsilon \red{\tilde{h}_1}(x,t) + O(\epsilon^2).
$
Here \red{$\tilde{h}_0$} is a dynamically evolving solution for \eqref{eqn:nondim} and $\red{\epsilon\tilde{h}_1}$ is a small perturbation of magnitude $\epsilon \ll 1$. Substituting this ansatz into \eqref{eqn:nondim} omitting terms of higher order in $\epsilon$, evaluating when the solution is locally constant, and omitting the diffusive terms gives
\begin{align}\label{eq:pert_convection}
    \red{\frac{\partial\tilde{h}_1 }{\partial t}} + f'(\tilde{h}_0)\red{\frac{\partial\tilde{h}_1 }{\partial x}} = 0.
\end{align}
From \eqref{eq:pert_convection} we may deduce the direction that the perturbations travel on either side of the shock. In the frame of the shock we note that the compressive and undercompressive shocks behave differently. For the compressive case \eqref{eq:lax} implies that perturbations will travel into the shock. In contrast, in the undercompressive case \eqref{eq:notlax} shows that perturbations travel through the shock.
This distinction in perturbation behavior is again illustrated in the characteristic plots in  FIG.~\ref{fig:characteristics}.
As in the undercompressive regime, perturbations travel down and away from the shock, the shock is stable to perturbations unlike the compressive case. 
We use this criteria as a signature to identify undercompressive shocks emerging from the meniscus.

\begin{figure}
\centering
 \includegraphics[width=0.55\linewidth, height=5.5cm]{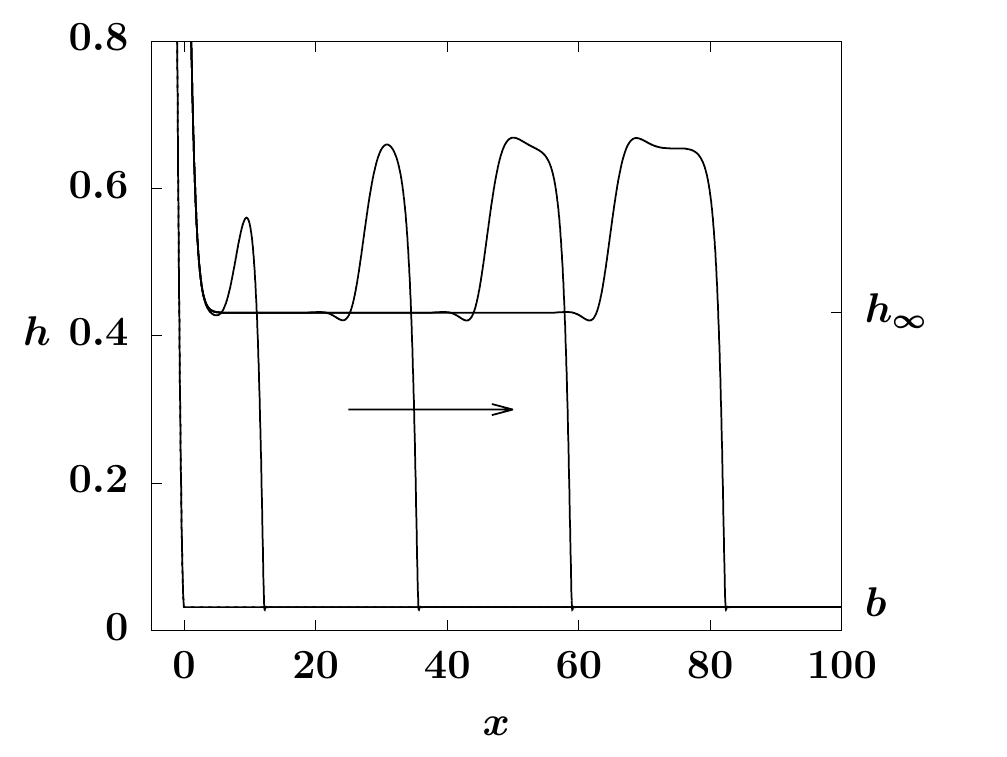}
  \includegraphics[width=0.44\linewidth,height=5.5cm]{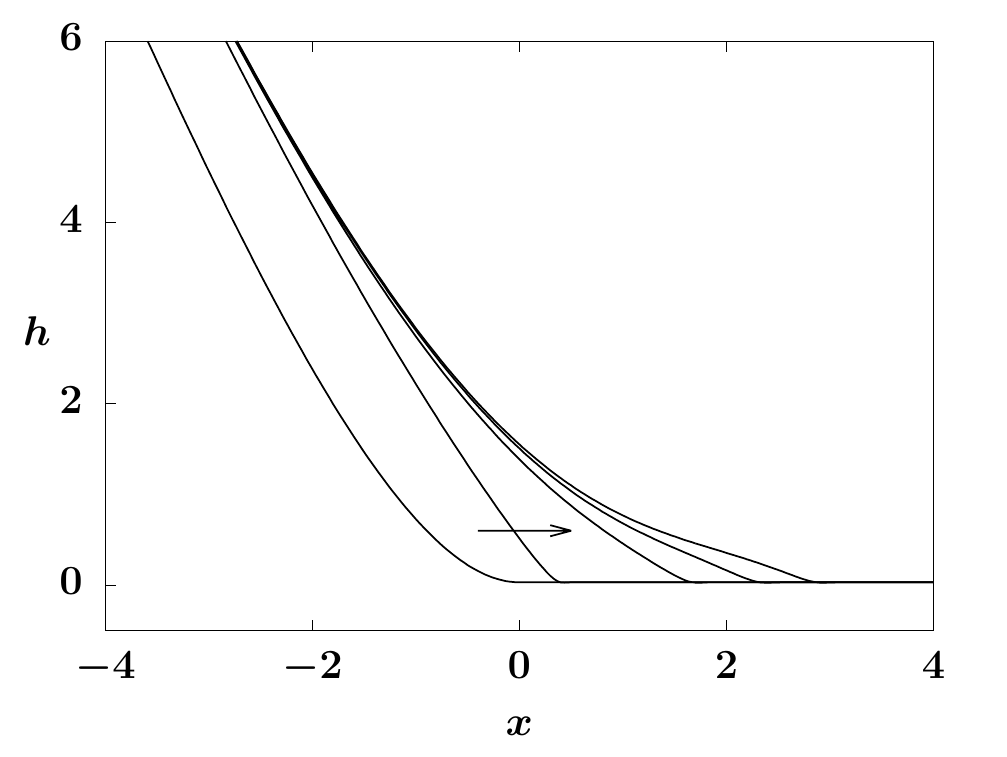}
  \caption{(left) Numerical simulation of \eqref{eqn:nondim} with parameters extracted from experiment (A) \cite{fournier1992tears}. The simulation illustrates a compressive-undercompressive shock pair in the long time with $b = 0.0317$ and $h_\infty = 0.43$.
  (right) Consecutive time steps $\sim 0.1  (s)$ starting from the meniscus initial condition. }
\label{fig:cazabat}
 
\end{figure}

\section{Meniscus-driven film climbing and nonclassical shocks}
\label{sec:shocktheory}

In this section we review prior published experimental results for this problem in which the film climbs onto a dry surface.  In this case, the meniscus controls the initial thickness of the film as it climbs, and we can solve equation \eqref{eqn:nondim} with the meniscus boundary condition, depending on the inclination angle.  This same model and boundary condition were already studied in \cite{munch2006interaction}, however the authors did not consider it in the context of the tears of wine problem. 

To model the dynamics of a spontaneous wine film climbing in a static glass, we approximate the boundary condition of equation \eqref{eqn:nondim} using a meniscus of fixed angle for the left boundary and a precursor pre-wetted layer for the right boundary following \cite{munch2006interaction}. 
The surface of wine in the bulk of the glass is horizontal and meets the thin film at a meniscus angle $\alpha$. This is expressed as a boundary condition describing the slope of the thin film with the glass,
${\partial h}/{\partial x} = \tan{\alpha}$.
In the non-dimensional settings, this gives 
$
\partial{h}/{\partial x} = -D^{-1}, 
$
and yields the far-field boundary condition
\begin{subequations}
\begin{equation}
h \to -x/D \qquad \text{ for } x \to -\infty.
\label{eq:lbc}
\end{equation}
For the thin precursor layer on the side of the glass 
we apply the boundary condition,
\begin{equation}
h \to b \qquad \text{ for } x \to \infty,
\label{eq:rbc}
\end{equation}
\label{eq:meniscus_bc}
\end{subequations}
where $b > 0$ is the precursor thickness.\red{ 
The precursor layer on the right boundary is commonly used as a replacement for more complicated contact line models  \cite{bertozzi1997linear}, and captures the relevant length scale at the contact line. This model alleviates complications that arise with a moving contact line in numerical simulations.}

Typical solutions of the PDE \eqref{eqn:nondim} subject to boundary conditions \eqref{eq:meniscus_bc} consist of two parts, the meniscus profile and the advancing front (see FIG.~\ref{fig:cazabat} (left)). 
The meniscus structure is a stationary solution of \eqref{eqn:nondim} satisfying the far--field boundary condition \eqref{eq:lbc} as $x \to -\infty$, and selects a flat state of thickness $h_{\infty} > b$ as the solution advances (see e.g. FIG.~ \ref{fig:cazabat}).  
FIG.~\ref{fig:cazabat} (right) shows that a stable meniscus solution is achieved in the numerical simulation of \eqref{eqn:nondim} starting from the initial data \eqref{ic},
\begin{equation}
    h_0(x) = 
    \begin{cases}
    D^{-3/2}(\exp(D^{1/2}x)-D^{1/2}x-1)+b &\text{ for }x\le 0,\\
    b & \text{ for }x > 0.\\
    \end{cases}
    \label{ic}
\end{equation}
This initial condition is a smoothed version of the piecewise linear function that captures the meniscus angle \cite{munch2006interaction}.
Starting from $h(x,\red{t =}0) = h_0(x)$, the simulation uses a standard finite difference spatial discretization and a backward implicit time-stepping scheme. The spatial derivatives are discretized using upwind scheme with respect to the flux $f(h)$, and central finite differences for the second and fourth derivative terms.

Away from the meniscus near the apparent moving contact line, the advancing front is given by a traveling wave that connects the left constant state $h_{\infty}$ and the right thin precursor layer $b$. Substituting the traveling wave ansatz $h(x,t) = h(\xi), \xi = x-st$ into \eqref{eqn:nondim}, and using the far field boundary condition \eqref{eq:rbc}, we get a third-order ODE that determines the advancing front profile 
\begin{subequations}
\begin{equation}
-s(h-b)+(f(h)-f(b))+h^3h'''-Dh^3h'=0,
\label{eq:tws_ode}
\end{equation}
subject to the far-field boundary conditions
\begin{equation}
    h \to h_{\infty} \,\text{ for } \xi \to -\infty,\qquad \,
    h \to b \quad \text{ for } \xi \to \infty,
    \label{eq:twsbc}
\end{equation}
\label{eq:tws}
\end{subequations}
where $' \equiv d/d\xi$. 
One can have zero, one, or multiple traveling waves depending on the values of the left and right states.  This is quite different from the case where the shock is smoothed by ordinary diffusion.  Surface tension results in a higher order equation with a complicated solution space \cite{bertozzi1999undercompressive,munch2006interaction}.  

To match front dynamics with different experiments, one can perform direct PDE simulations of model \eqref{eqn:nondim} using the meniscus boundary conditions \eqref{eq:meniscus_bc}.
For instance, in FIG.~\ref{fig:cazabat} (left), corresponding to the experiment in \cite{fournier1992tears}, the meniscus dynamics with given $(D,b) = (0.353,0.0317)$ selects a flat state thickness $h_{\infty}>b$, and the advancing front consists of two different types of shocks: a compressive shock in the rear and an undercompressive shock at the front of the film. 
More generally, distinct solution behaviors involving various types of meniscus profiles and advancing fronts can emerge with $(D,b)$ in different parameter regions;
this have been extensively studied in \cite{munch2006interaction}.

\begin{figure}
 \centering
  \begin{minipage}[b]{0.59\textwidth}
   \centering
    \begin{tabular}{|c|c|c|}
        \hline
    \textbf{Shock types} & \textbf{Stability} & \textbf{Figures}\\ \hline
        Compressive shock   & unstable &  FIG.~\ref{fig:cazabat} (right) \\
      \hline
      \makecell{  Compressive-\\
      undercompressive \\
      double shock}
  
 & \makecell{unstable\\
 stable}
 & 
 \makecell{ FIG.~\ref{fig:cazabat}~(left),\\ FIG.~\ref{fig:vueill}~(left)
}
 \\
        \hline
        \makecell{
      Rarefaction-\\
     undercomperssive shock  }
     & \makecell{
     stable}
     & FIG. \ref{fig:vueill} (right)\\
         \hline
         \makecell{
         Rarefaction-\\
      Reverse--undercompressive \\
      shock  }
     & \makecell{
  unstable}
     & Section \ref{sec:RUC}\\
    \hline
  \end{tabular}
      
  \vspace{0.2in}
    \label{tab:shock_type}
    \end{minipage}
\hfill
  \begin{minipage}[b]{0.39\textwidth}
    \includegraphics[width =5.5cm]{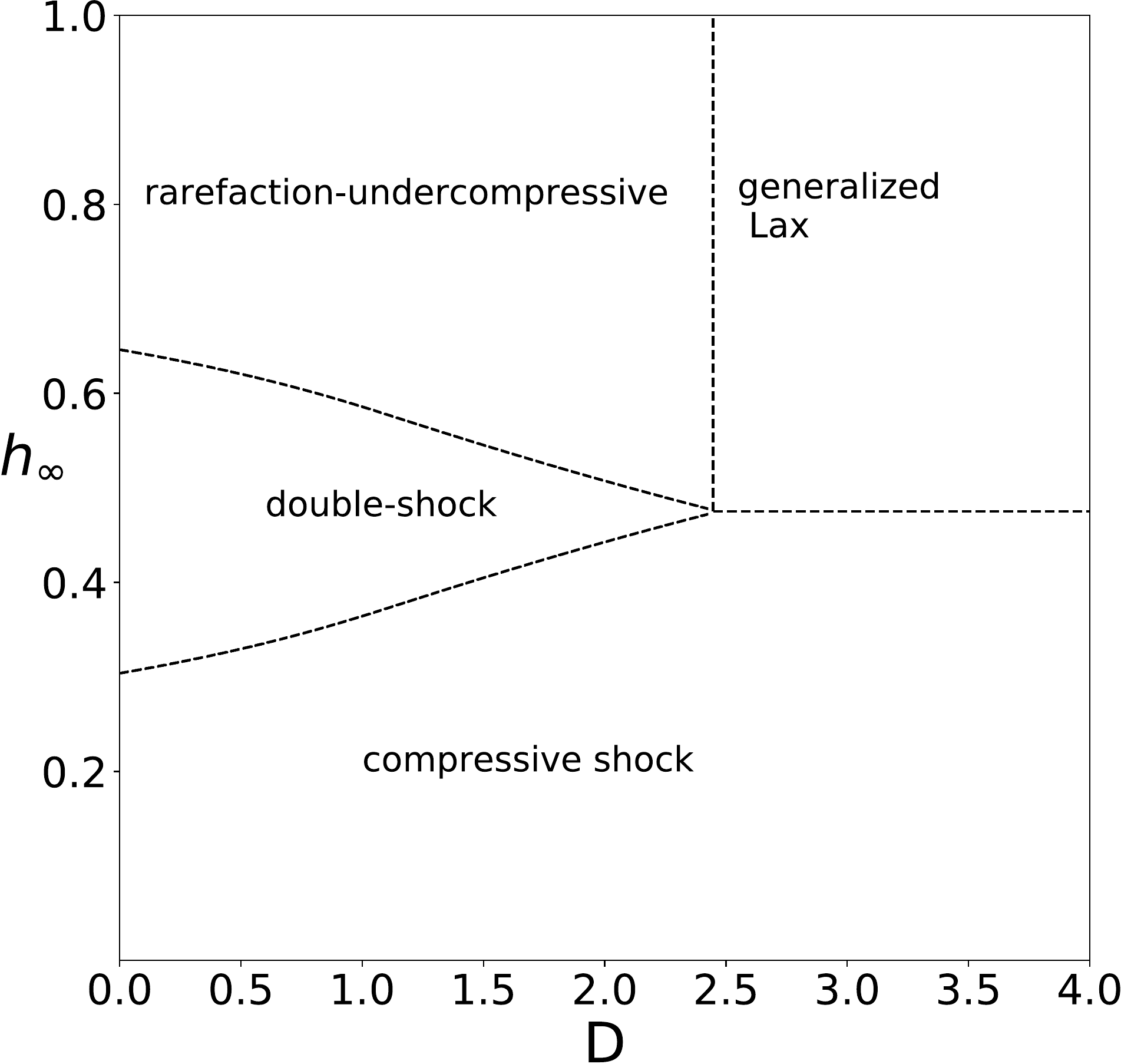}
    \end{minipage}
  \caption{(left) Shock types of the hydro-dynamic model \eqref{eqn:nondim} discussed in the present study; (right) A shock bifurcation diagram parametrized by $(h_{\infty},D)$ pairs obtained by numerically solving the ODE \eqref{eq:tws_ode} subject to the boundary condition \eqref{eq:twsbc} for $b = 0.0353$.}
  \label{fig:shockTypes}
\end{figure}

Alternatively, for given values of $(D, b, h_{\infty})$, one may also use traveling wave solutions satisfying the ODE \eqref{eq:tws} to identify the features of the advancing front \cite{munch1999rarefaction,munch2006interaction}. Here the thickness of the left state $h_{\infty}$ can either be measured experimentally or calculated numerically based on the meniscus dynamics.
Instead of revisiting the full dynamics of the meniscus-driven film climbing problem, we briefly review possible shock scenarios characterized by the traveling wave solutions in the context of tears of wine. 
For a fixed dimensionless $b = 0.0353$, corresponding to the precursor thickness in an experiment from \cite{vuilleumier1995tears}, FIG.~\ref{fig:shockTypes} (right) summarizes four possible shock scenarios parametrized by $h_{\infty}$ and $D$. 
This bifurcation diagram is numerically obtained by solving the ODE \eqref{eq:tws_ode} using the asymptotic boundary condition method \cite{golovin2001effect}, and is similar to the one studied in \cite{munch2000shock} for shock transitions in Marangoni gravity-driven thin films.

Four plausible shock structures for $(h_{\infty}, D)$ in different parameter regions are depicted in FIG.~\ref{fig:shockTypes} (right):
(1) a single compressive shock, 
(2) a separating double shock pair involving a leading undercompressive wave and a trailing compressive wave (see FIG.~\ref{fig:cazabat} (left)),
(3) a rarefaction-undercompressive shock structure, and (4) a generalized Lax shock. 
The bifurcation diagram shows that for small values of $D$, as in most tears of wine experiments from the literature, only shock wave structures of type (1), (2), and (3) can exist. 
We will discuss these cases using experimental data in the next section.
We also present the stability properties of these shocks with respect to transverse perturbations, and point to their corresponding figures in  FIG.~\ref{fig:shockTypes} (left). In particular, the compressive shock is linearly unstable to transverse perturbations which plays an important role in developing later-stage fingering patterns. In contrast, in both the compressive--undercompressive shock pair and the rarefaction--undercompressive shock, the leading undercompressive front is stable and prevents fingering from happening in the contact line \cite{bertozzi1999undercompressive, bertozzi1998contact,bowen2005nonlinear}.

Another type of shock, reverse--undercompressive shock, is also observed in the study of tears of wine dynamics after a glass swirling. Modified initial and boundary conditions will be used to characterize this scenario. This is documented in FIG.~\ref{fig:shockTypes} (left), and we will discuss this case in detail in section \ref{sec:RUC}.

\section{Experimental survey and simulations}
\label{sec:survey}
Now that we have a nonlinear model for the wetting behavior of the climbing film, we can compare it with experimental data in the prior literature. \red{ However, the behavior of undercompressive shocks depends very sensitively on the dimensionless parameter $b$.
Very few experiments study this in detail - one example being \cite{schneemilch2000shock} for thermally driven films which are easier to control. } 
Likewise $\tau$ can sometimes be hard to measure and it appears in the calculation of both $b$ and $D$, which are the dimensionless parameters needed to model the experimental data.
We analyze the effect of the uncertainty of these parameters here.

We consider the prior works: (A) the seminal ``Tears of Wine'' \cite{fournier1992tears} paper and (B) ``Tears of wine: the stationary state'' \cite{vuilleumier1995tears}. (A) presents several experiments from which we use the parameters corresponding to alcohol concentration $C = 70\%$. This experiment has the most detailed measurements and also shares some measurements with Vuiellemuier \textit{et al.}\cite{vuilleumier1995tears}.
For (B) we analyze two physical experiments: Experiment I, that follows the experimental settings of Figure 5b of their paper with a curvature-driven film and $C = 70 \%$, and experiment II that refers to Figure 5b of (B) and follows a gravity-driven regime with $C = 70\%$. We label the experiments as (BI) and (BII) and note that they correspond to the same physical setting with different assumptions when deriving the surface tension gradient $\tau$.

\begin{figure}
\centering
  \includegraphics[width=0.48\linewidth]{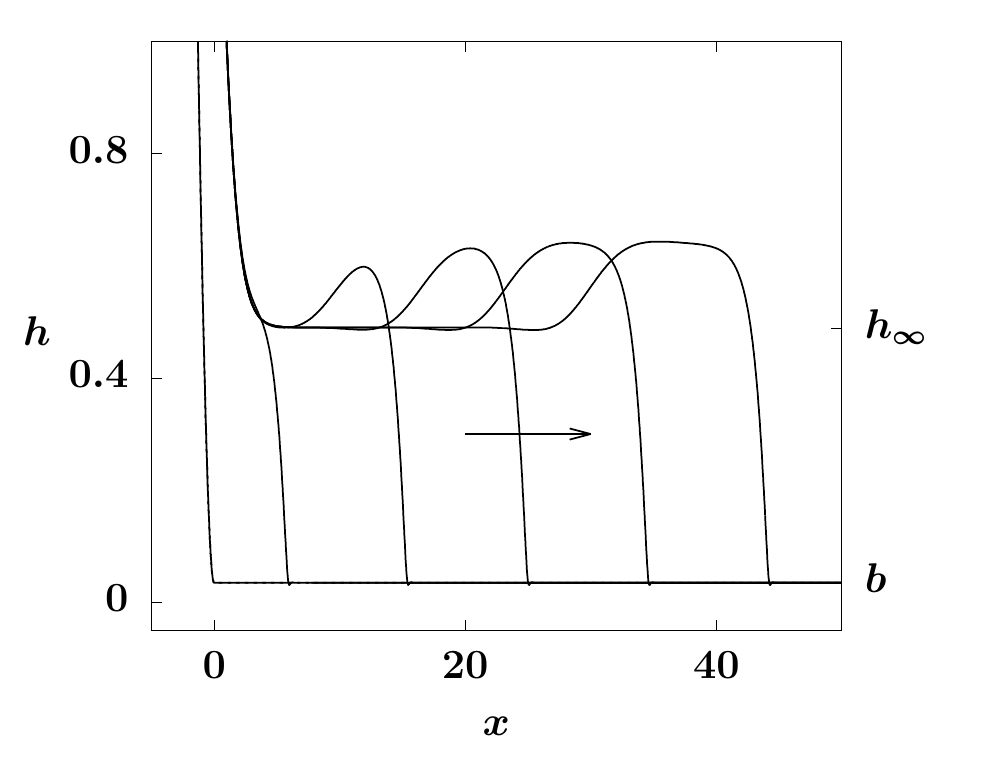}
 \includegraphics[width=0.48\linewidth]{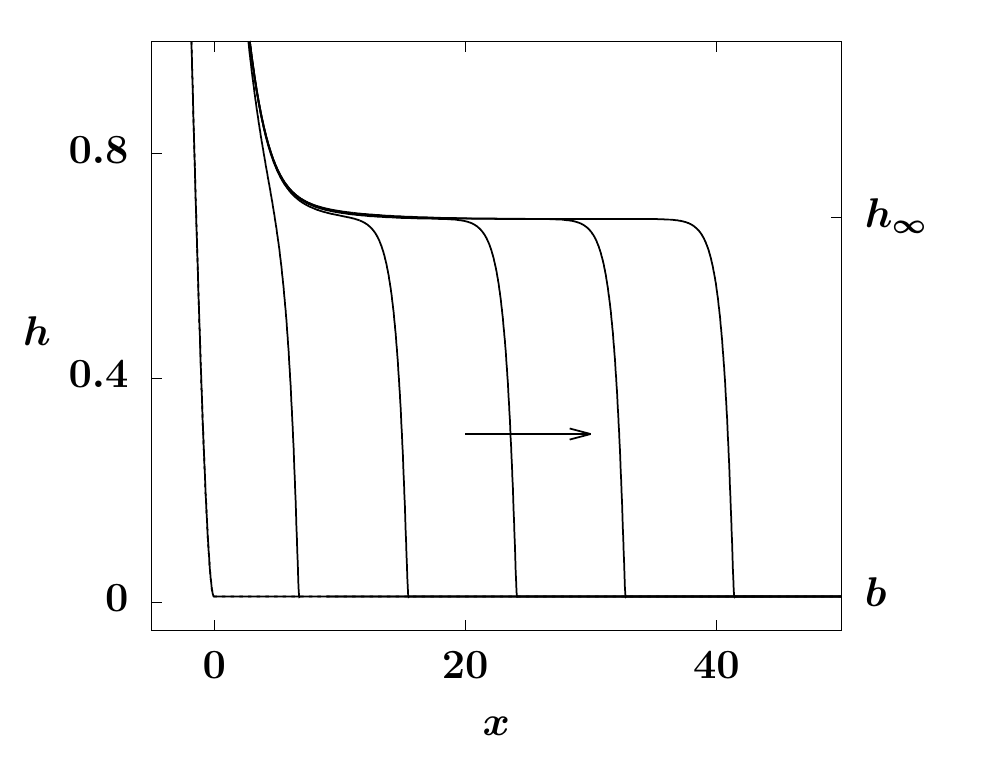}
    \caption{Numerical simulations of Vuiellemuier \textit{et al.}experiments (BI), (BII) for long times. (left) experiment (BI) exhibiting a less distinct compressive-undercompressive double shock cf. FIG.~\ref{fig:cazabat}. (right)  experiment (BII) exhibiting undercompressive shock.
    }
 \label{fig:vueill}
\end{figure}

In Appendix \ref{appendix:extended_survey} we provide a complete set of measurements for each experiment, as well as the dimensionless values $(D,b)$ needed for analysis. Using different $(D,b)$ values corresponding to each experiment we conduct a sequence of numerical simulations for equation \eqref{eqn:nondim}. The initial and boundary conditions are specified as in (\ref{eq:meniscus_bc} -- \ref{ic}).
In FIG. \ref{fig:vueill} we present numerical simulations for (BI) and (BII) and observe that despite identical physical settings the different values of $\tau$ lead to different shocks. In particular (BI) exhibits a compressive-undercompressive shock while (BII) has a single undercompressive shock front. 

\begin{figure}
\centering
 \includegraphics[width=7cm, height=5.2cm]{ExperimentFournierCazabat_D353e-3_b317e-4.pdf}
\includegraphics[width=7cm,height=5.2cm]{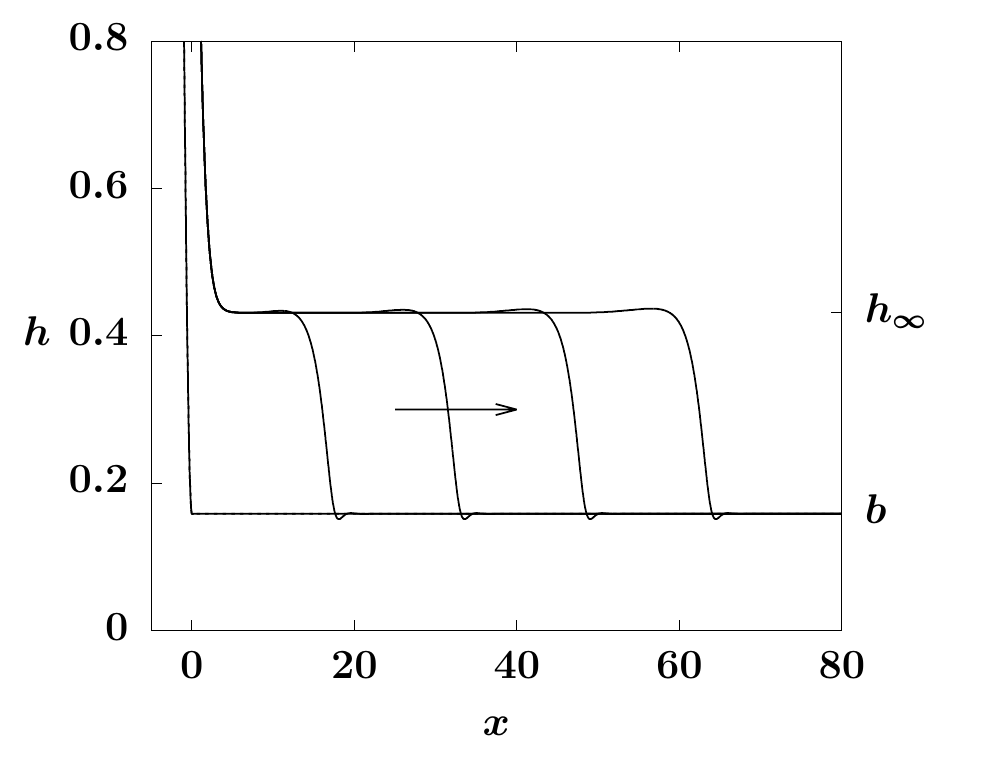}
\caption{A comparison of shock types affected by the precursor thickness $b$, showing that (left) $b = 0.0353$ results in a compressive-undercompressive shock, and (right) $b = 0.1585$ (corresponding to a hypothetical dimensional thickness $\bdim=10 \mu$m) leads to the formation of a compressive wave. The other parameters in the two simulations are identical to those in FIG.~\ref{fig:cazabat} and correspond to measurements from experiment (A).
} 
\label{fig:cazabat_meniscus}
 
\end{figure}

In addition to $\tau$, the precursor thickness $b$ is also of key importance to the dynamics of the advancing front \cite{bertozzi1998contact}. For example for the setting of (A) that leads to an advancing front with a compressive--undercompressive double shock, when the precursor thickness is increased from $b = 0.0353$ to $b = 0.1585$ the front transitions into a single compressive wave. We observe this change in behavior while the rest of the settings are fixed (see FIG.~\ref{fig:cazabat_meniscus}).

 \begin{figure}
\centering
\includegraphics[width=7.7cm,height=5.2cm]{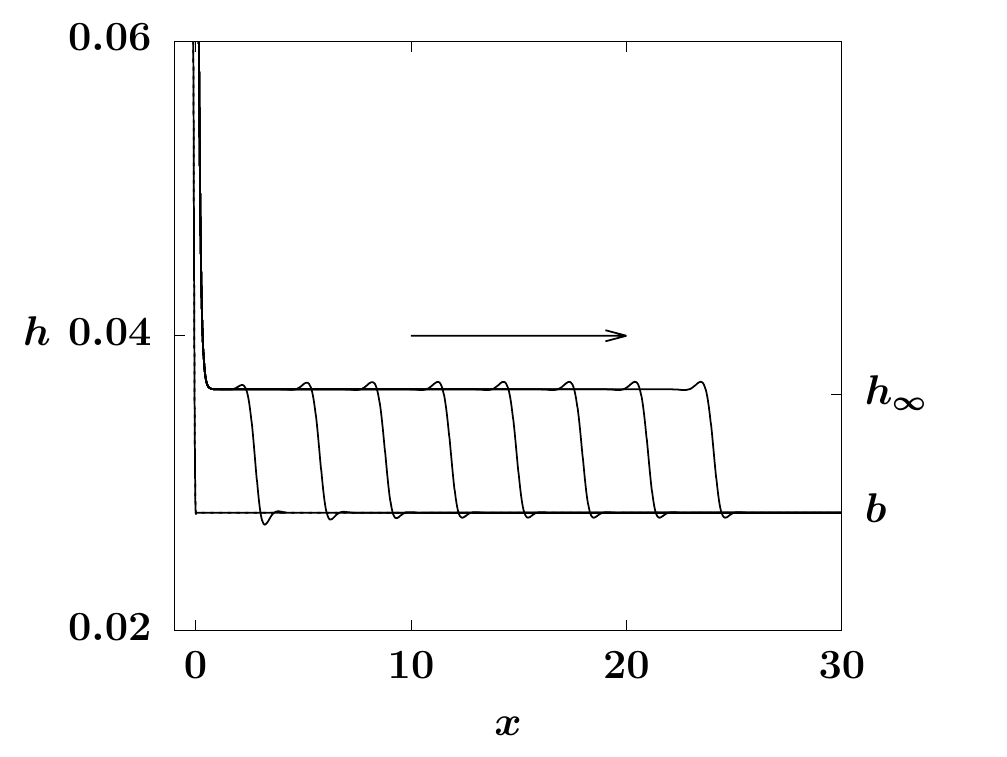}
\caption{The evolution of film height of \cite{venerus2015tears} (wine setting) with $b = 0.028$ (corresponding to a hypothetical dimensional thickness $\bdim=0.5\mu$m) showing the formation of a compressive wave.
The meniscus dynamics with $D = 0.0338$ yields a left state thickness $h_{\infty} = 0.036.$
} 
\label{fig:experimentDI}
\end{figure}

For other experiments in the literature \cite{hosoi2001evaporative, venerus2015tears} (see Appendix \ref{appendix:extended_survey} for complete listing) the authors did not report their data for the climb of the film. Therefore we cannot fully compare our theory against their experimental observations. 
In our experiments that match the high inclination angle and ethanol-water fraction of \cite{hosoi2001evaporative, venerus2015tears}, we do not observe easily reproducible film climbing.

\begin{figure}
\centering

\includegraphics[width=0.3 \textwidth]{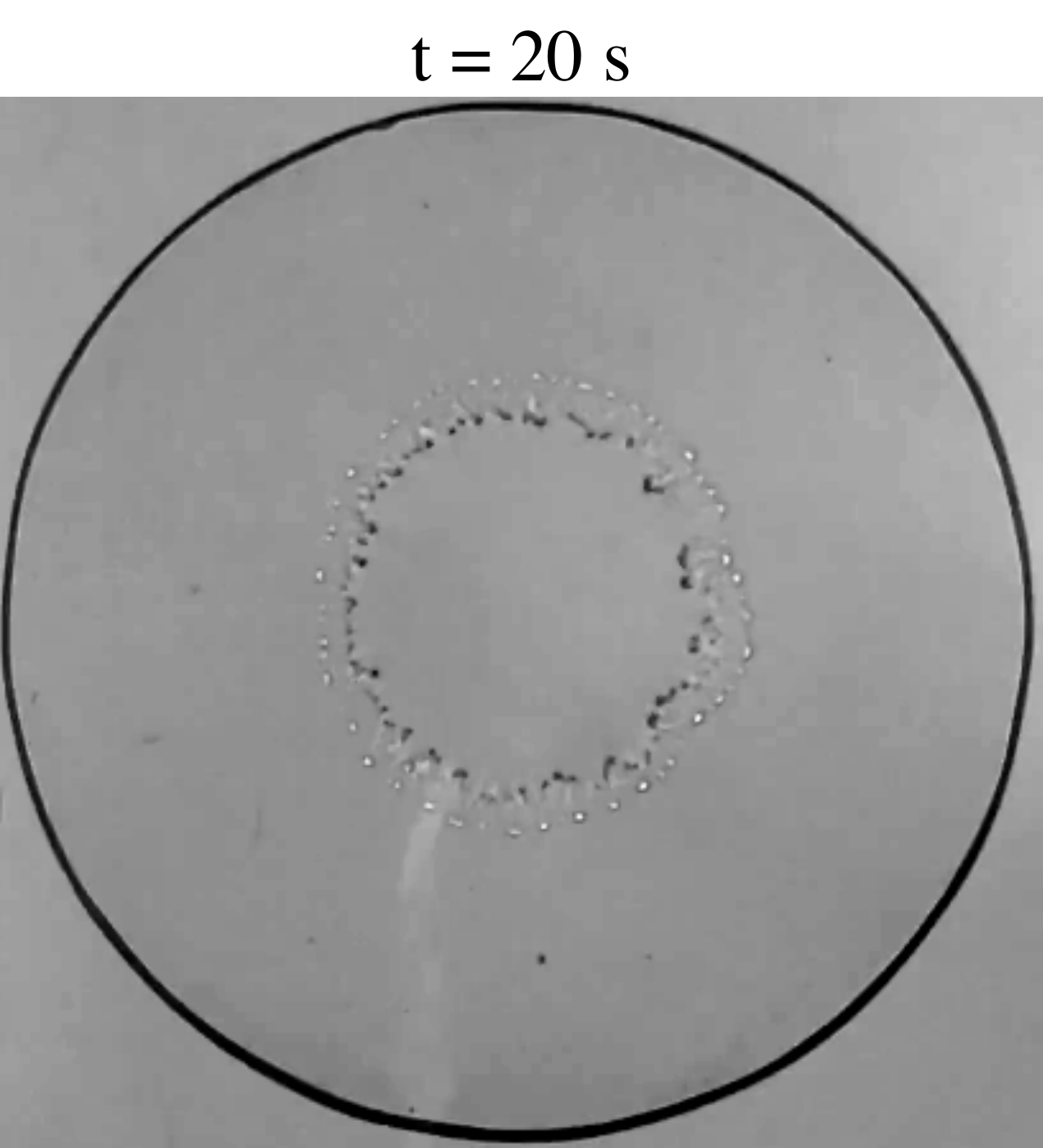}
\includegraphics[width=0.3 \textwidth]{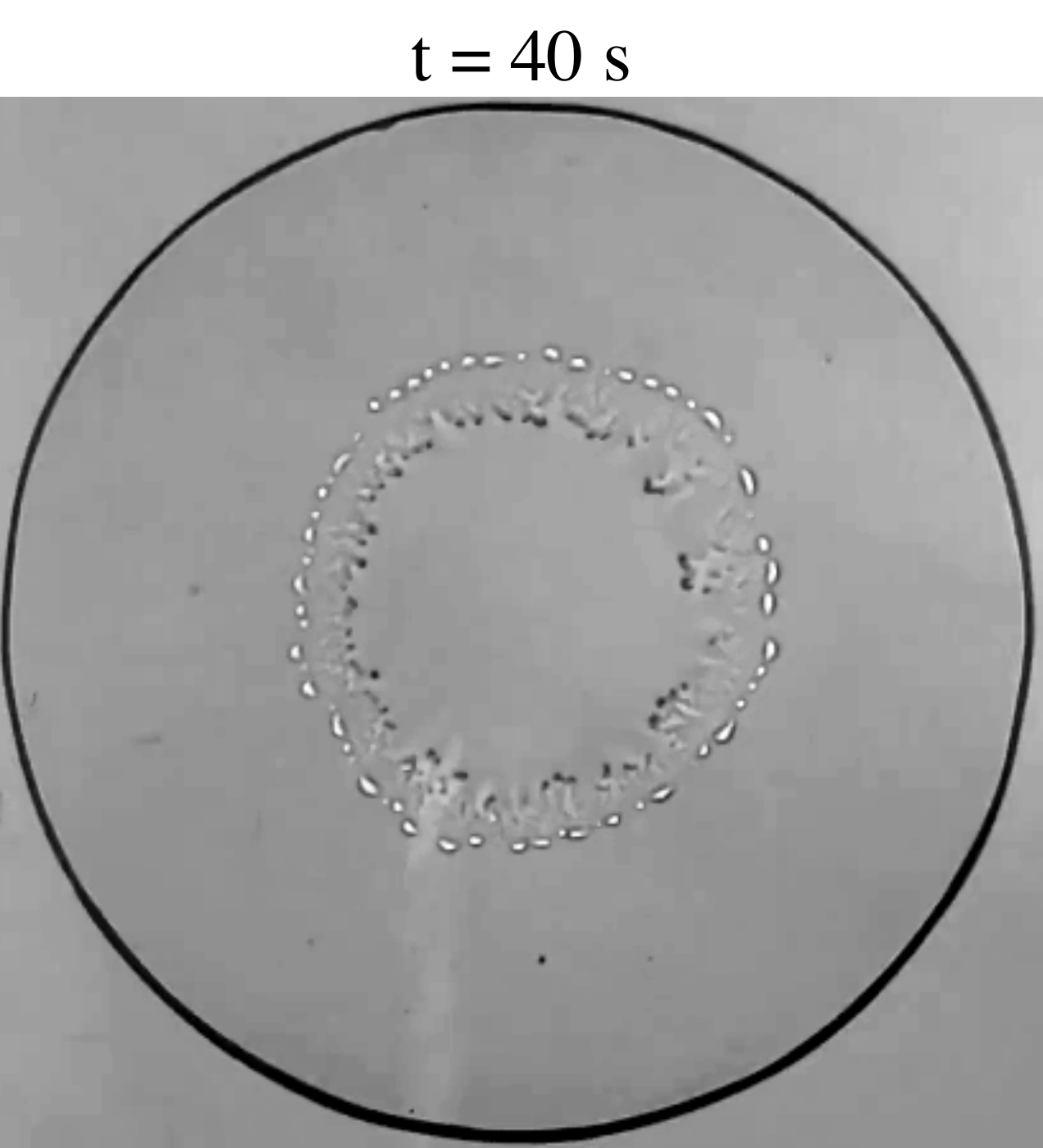}
\includegraphics[width=0.3 \textwidth]{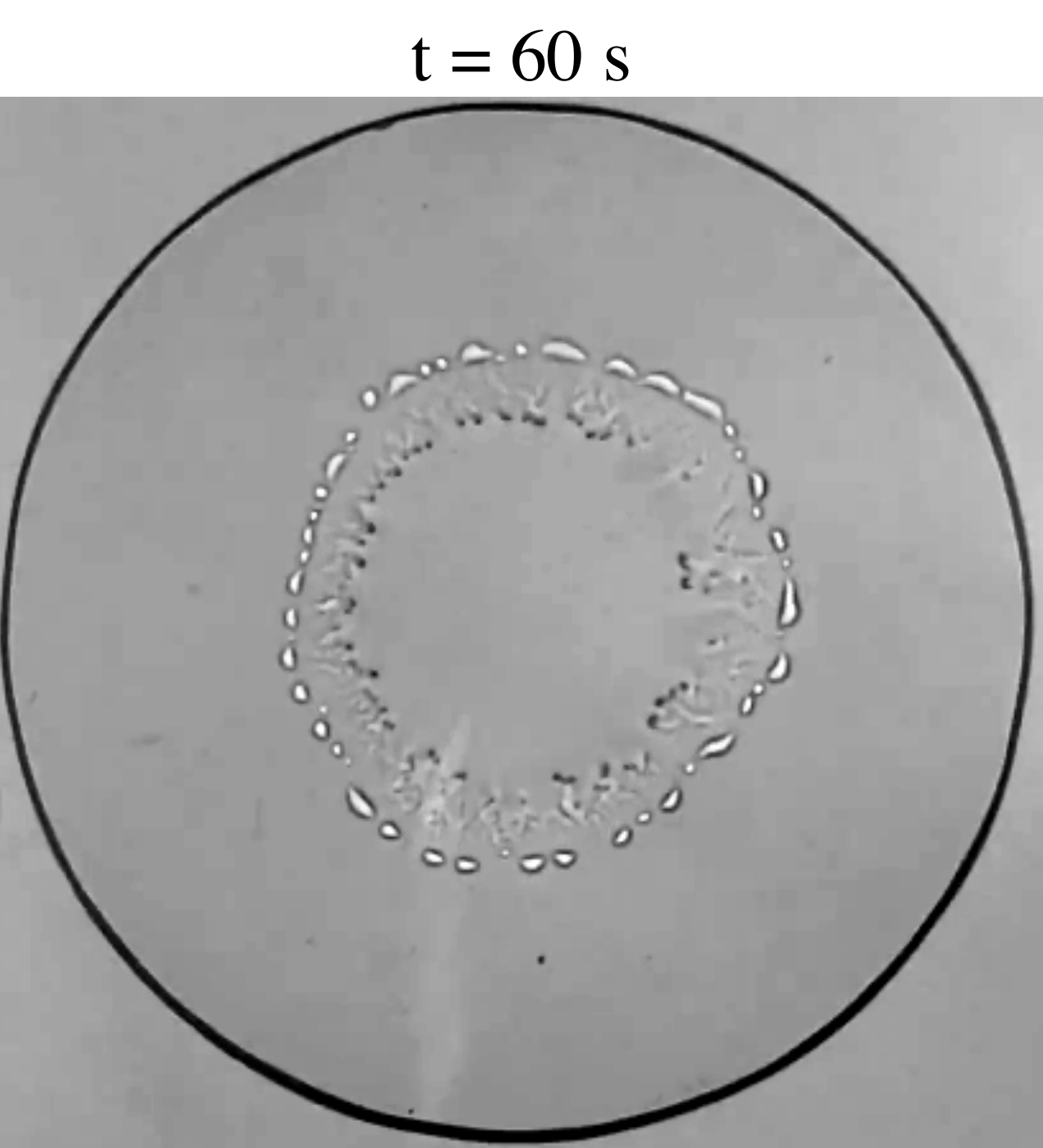}

\caption{Spontaneous climb images of ethanol-water mixture with ethanol concentration $C = 0.7 $ on a dry watch glass (no pre-swirl) of diameter 75 mm and angle ranging between $9 \degree <\alpha < 20 \degree $.} 
\label{fig:Ethanol_experiment}
\end{figure}

With a hypothetical thin precursor thickness, our simulations of \cite{hosoi2001evaporative, venerus2015tears} based on the meniscus-driven film dynamics predict a thin compressive advancing front. In FIG.~\ref{fig:experimentDI} we present the evolution of a thin film climbing out of the meniscus using the dimensionless parameter $D=0.0338$ that corresponds to Table 3, Figure 10 of  \cite{hosoi2001evaporative}. A small precursor thickness $b=0.028$ (corresponding to a dimensional thickness of $\bdim=0.5\mu$m) is used to approximate the dry substrate.
For our experiment, in FIG. \ref{fig:Ethanol_experiment}, we show a spontaneous climb that is similar to experiment (A). The settings \red{of our experiments} are of a watch glass of diameter 75 mm and angle  $9 \degree <\alpha < 20 \degree $ \red{(due to curvature of the watch glass)}. For this high alcohol concentration and inclination, the climbing film on the dry substrate does exhibit a spontaneous climb.

In our model, different settings lead to different shock structures. This is in contrast with the previous literature, where a model with a surface tension gradient and tangential gravity is used \red{and only a single type of shock emerges}. Without the competition between the surface tension gradient $\tau$ and the tangential component of gravity, we only observe classical compressive shocks. When incorporating both gravity and surface tension as in \eqref{eqn:nondim}, different physical parameters (in particular the substrate wetting thickness $b$) lead to qualitatively different shocks.  In Appendix \ref{appendix:extended_survey} we present tables of the prior works with some photographs from the experiments. 
More work is needed to better understand quantitatively how shocks behave in tears of wine on a dry surface.  Going forward here, we show that in the case of a \red{surface coated by swirling}, one can obtain very reproducible shock profiles, which our theory suggests are reverse undercompressive shocks.\red{ We distinguish pre-swirling from the precursor discussed in Sections \ref{sec:level2} -- \ref{sec:curvatureEffects}, noting that in the preswirled regime, the right boundary condition maintains a thicker fluid film}.
This is further discussed in section~\ref{sec:RUC}.  In the next section we present a model for a conical shaped substrate (as in our experiments) rather than a flat substrate.  We show that it results in minor modifications to the behavior.  The flat surface case is important because the model reduces to a regular scalar conservation law for which there is a well-developed shock theory.

\section{Conical shaped substrate}
\label{sec:curvatureEffects}

So far we have assumed negligible curvature effects of the substrate. In this section we investigate the substrate curvature effects on shock formation. For simplicity, we consider an axisymmetric thin fluid film climbing up the surface of a conical-shaped cocktail glass of inclination angle $\alpha$ (see FIG.~\ref{fig:towlabel} (left)). 
In the long-wave limit the balance of normal stresses at the free surface $z = h(x,t)$  yields the leading-order equation 
\begin{equation}
    p = \frac{A}{x+x_0} - \frac{\partial^2 h}{\partial x^2},
    \label{pressure}
\end{equation}
where $p$ is the dynamic pressure, the term $A/{(x+x_0)}$ represents the azimuthal curvature of the conical substrate, $x_0>0$ measures the distance between the surface of the wine reservoir/bulk and the vertex of the cone, and the non-dimensional parameter $A$ is given by $A = {X}/{(H\cot{\alpha})}$ for the length-scales $X$ and $H$ defined in \eqref{scaling}. 
Using a PDE derived in \cite{roy2002lubrication} and studied in \cite{greer2006fourth} for the dynamics of thin films driven by gravity and surface tension on a curved substrate, we write
the non-dimensional governing equation for the film thickness $h(x,t)$ as
\begin{equation}
\frac{\partial \zeta}{\partial t} + \frac{\partial}{\partial x}\left(h^2 - h^2\zeta\right) = \frac{\partial}{\partial x}\left(h^2\zeta\frac{\partial p}{\partial x}\right)+D\frac{\partial}{\partial x}\left(h^3\frac{\partial h}{\partial x}\right), \qquad x \ge 0 ,
\label{modelcurvature_full}
\end{equation}
where $\zeta$ represents the amount of fluid above a surface patch and is approximated by
\begin{equation}
\zeta = h - \frac{\beta h^2}{x+x_0},
\end{equation}
where the non-dimensional quantity
$\beta = {H}/({2X\cot{\alpha}})$
arises from the principle curvature of the substrate.
This model characterizes the joint effects of substrate curvature, constant surface tension gradient, and both normal and tangential components of the gravity.
\begin{figure}
\centering
\includegraphics[width=7cm]{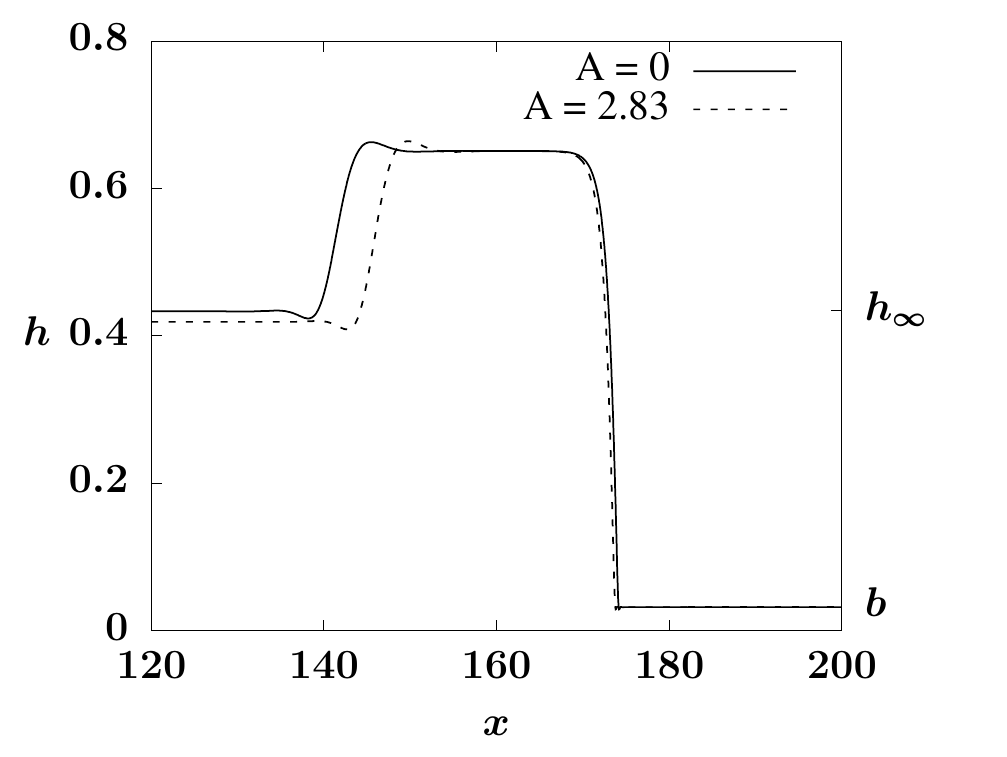}
\caption{Long-time shock profiles of \eqref{modelcurvature} for $A=0$ (no curvature effects) and $A=2.83$, $x_0 = 5$ (with curvature effects) at $t = 700$. 
Other system parameters are $b = 0.0317$, $D = 0.353$ corresponding to experiment (A).} 
\label{fig:curvatureLargeAngle}
\end{figure}
For typical tears of wine experiments we have $H/X \ll 1$ and $\beta \ll 1$, therefore we approximate $\zeta$ by $h$, and rewrite equation \eqref{modelcurvature_full} using \eqref{pressure}  by
\begin{equation}
h_t + (h^2 - h^3)_x =  \displaystyle{-\left[h^3\left(-\frac{A}{(x+x_0)^2}+h_{xxx}\right)\right]_x } +D\left(h^3h_x\right)_x.
\label{modelcurvature}
\end{equation}
Here it is important to have $x_0 > 0$ to avoid the shape singularity at the vertex.
In the limit $x\to \infty$, the azimuthal curvature term is dropped and the model \eqref{modelcurvature} reduces to \eqref{eqn:nondim}.

Using experimental parameters in experiment (A) with a small inclination angle $\alpha = 9\degree$, we plot in FIG. \ref{fig:curvatureLargeAngle} the comparison of long-time shock profiles without curvature effects $(A = 0)$ and with finite curvature effects $(A = 2.83,~x_0 = 5)$. 
It shows that incorporating the substrate curvature effects lowers the thickness of the left constant state $h_{\infty}$, and makes the separation of
the leading undercompressive wave and the trailing compressive wave in the double shock pair less pronounced.
Based on the theory for shock transitions in model \eqref{eqn:nondim} (or equivalently \eqref{modelcurvature} with $A = 0$) developed in \cite{bertozzi1999undercompressive,munch2000shock}, for fixed $D$ and $b$ values, decreasing the value of $h_{\infty}$ can push the solution out of the double shock regime and into the compressive regime (see FIG.~\ref{fig:shockTypes} (right)). This is consistent with our observation in FIG.~\ref{fig:curvatureLargeAngle} with finite curvature effects ($A > 0$), where the less pronounced separation of fronts caused by the decreased $h_{\infty}$ suggests a transition to compressive waves.

While the model \eqref{modelcurvature} is limited to the dynamics on a conical shaped substrate, a generalized nonlinear model incorporating the substrate geometry of wine glasses can be obtained by using a different functional term for the azimuthal curvature term.
More complicated curvature-induced shock transitions are expected to occur, and we refer the readers to the work of Roys \textit{et al.}\cite{roy2002lubrication} and Greer \textit{et al.}\cite{greer2006fourth} for a detailed discussion of the modeling and numerical methods of lubrication models on a curved substrate. 

\section{Reverse undercompressive shocks on a preswirled substrate}
\label{sec:RUC}

\begin{figure}
\centering 
\includegraphics[width=3.2cm]{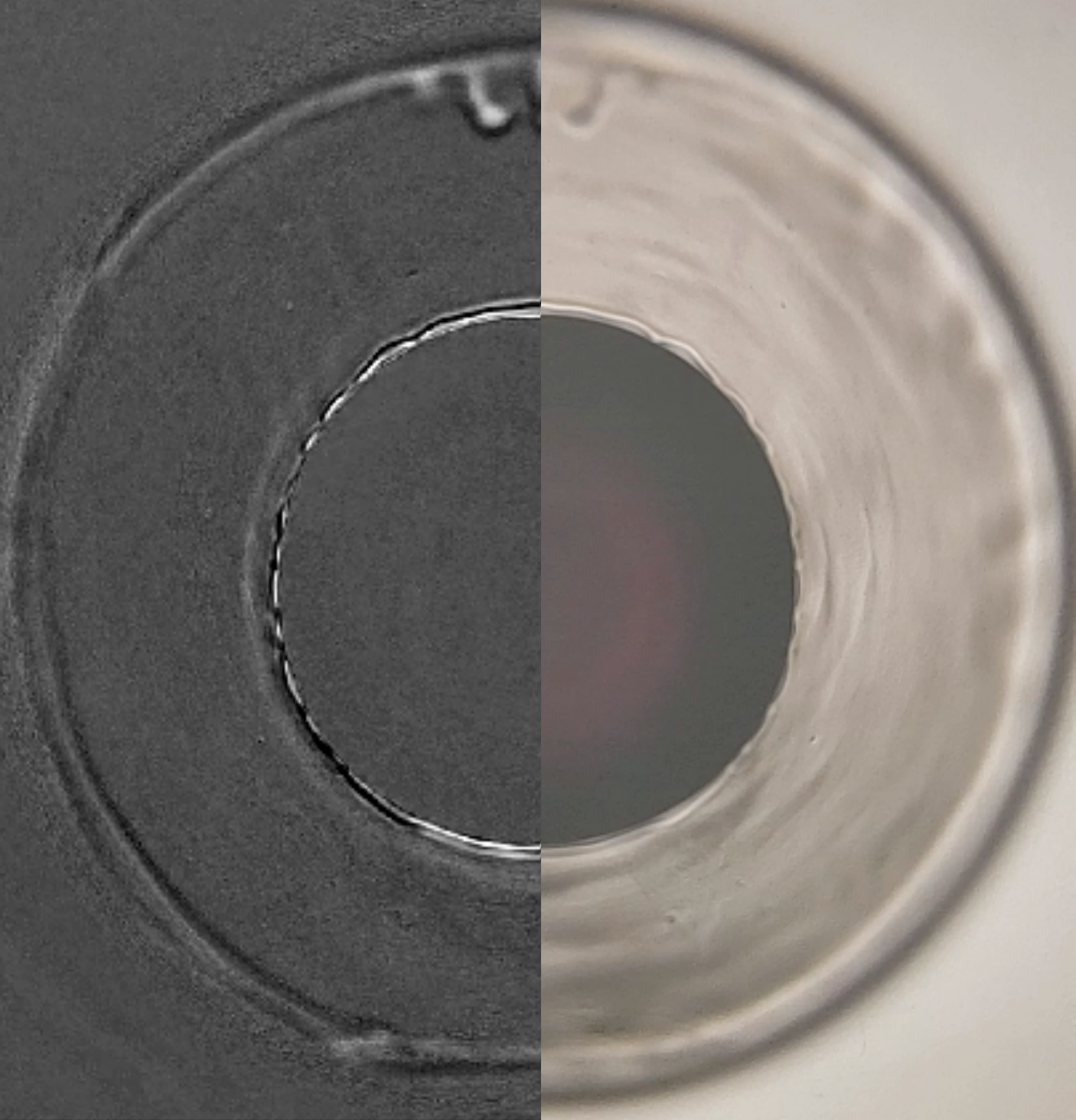}
\includegraphics[width=3.2cm]{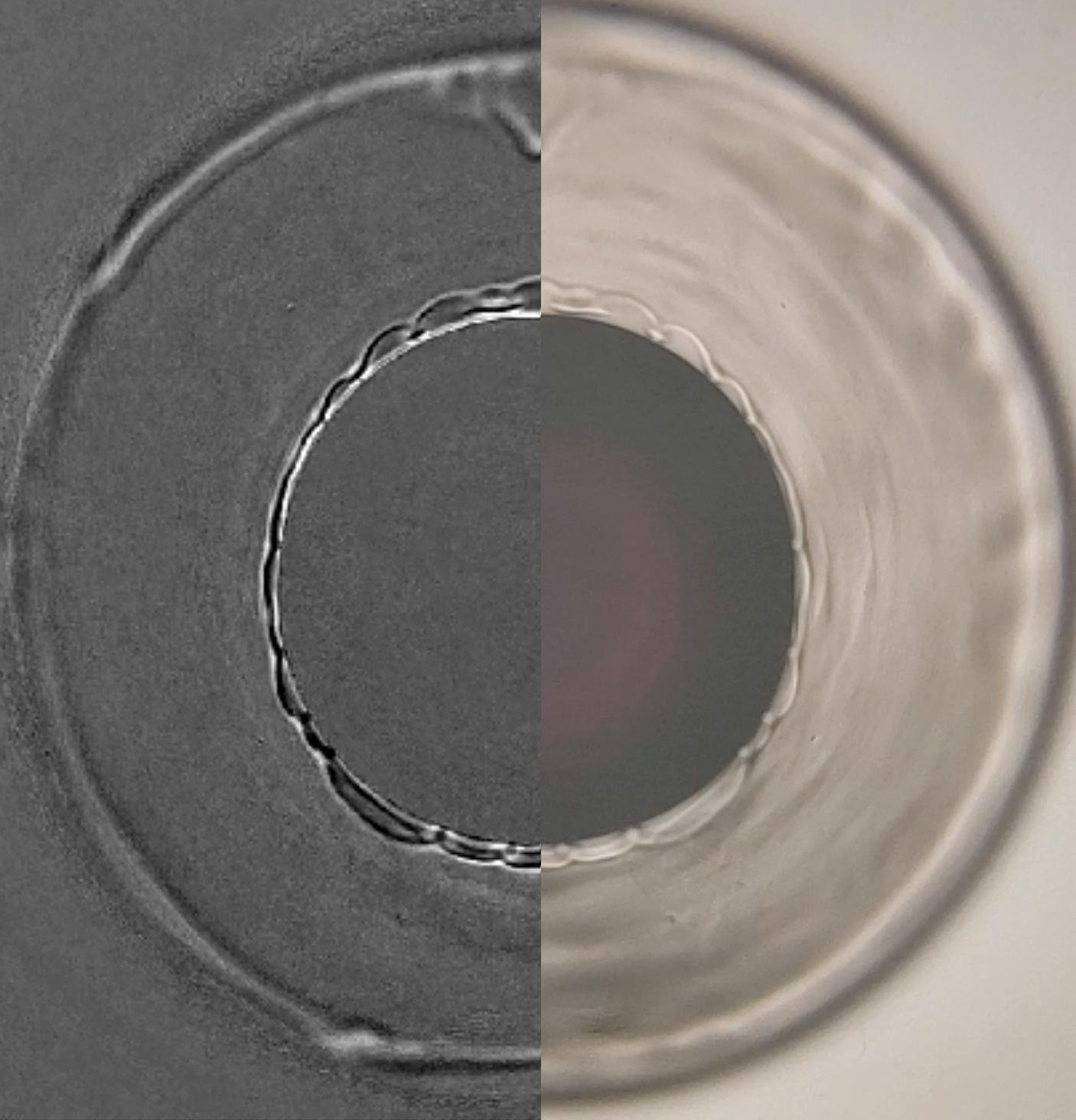}
\includegraphics[width=3.2cm]{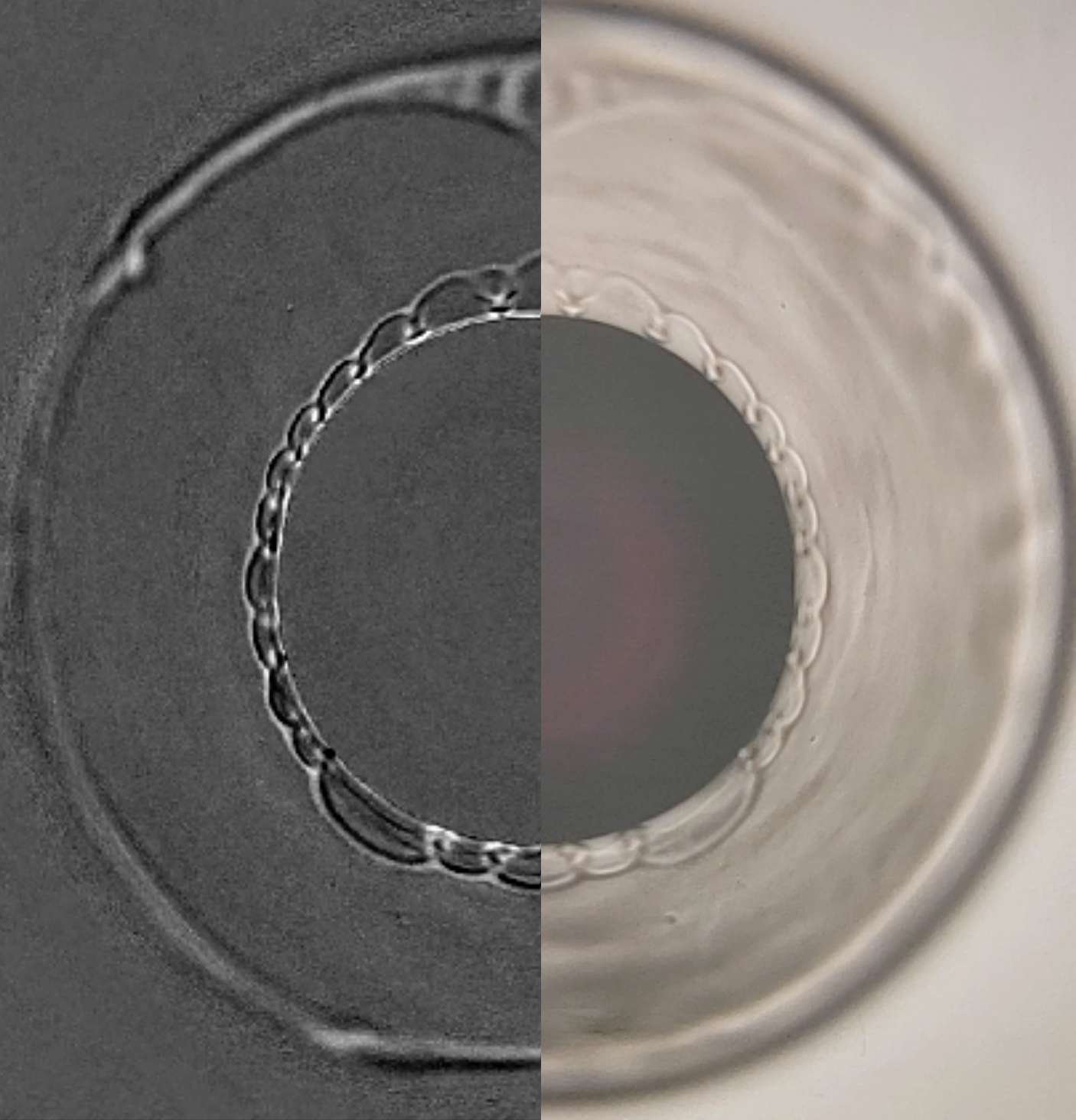}
\includegraphics[width=3.2cm]{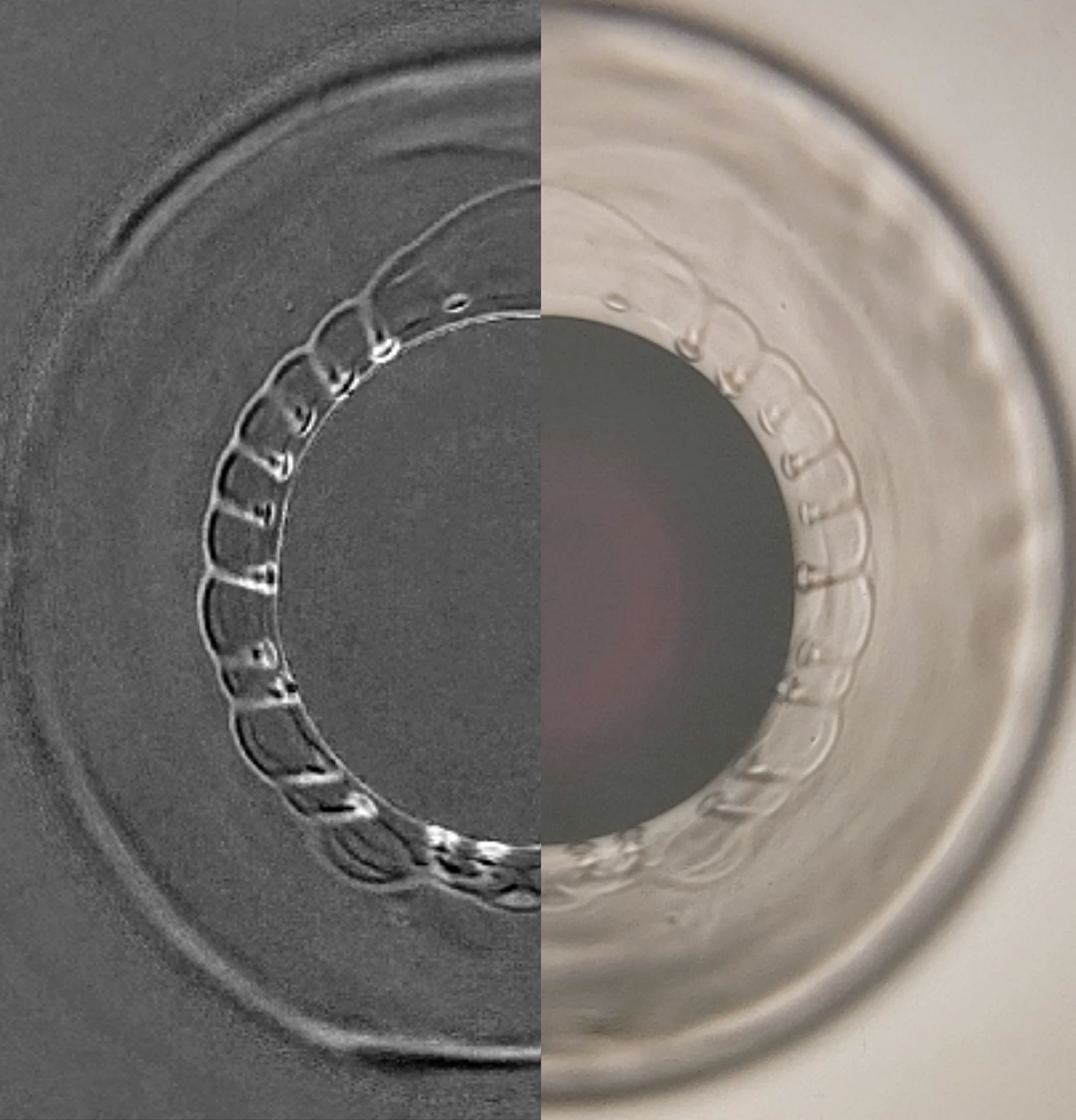}
\caption{Top view images of tears of wine experiment at $t=0, 5, 10, 20$ s in a stemless Martini glass (conical substrate) using $18\%$ alcohol by volume Port wine. Swirling the wine around the glass creates a reverse front that forms out of the meniscus, advances up the glass and destabilizes into wine tears.} 
\label{fig:RUC_experiment}
\end{figure}

\begin{figure}
\centering
\includegraphics[width=0.5\textwidth]{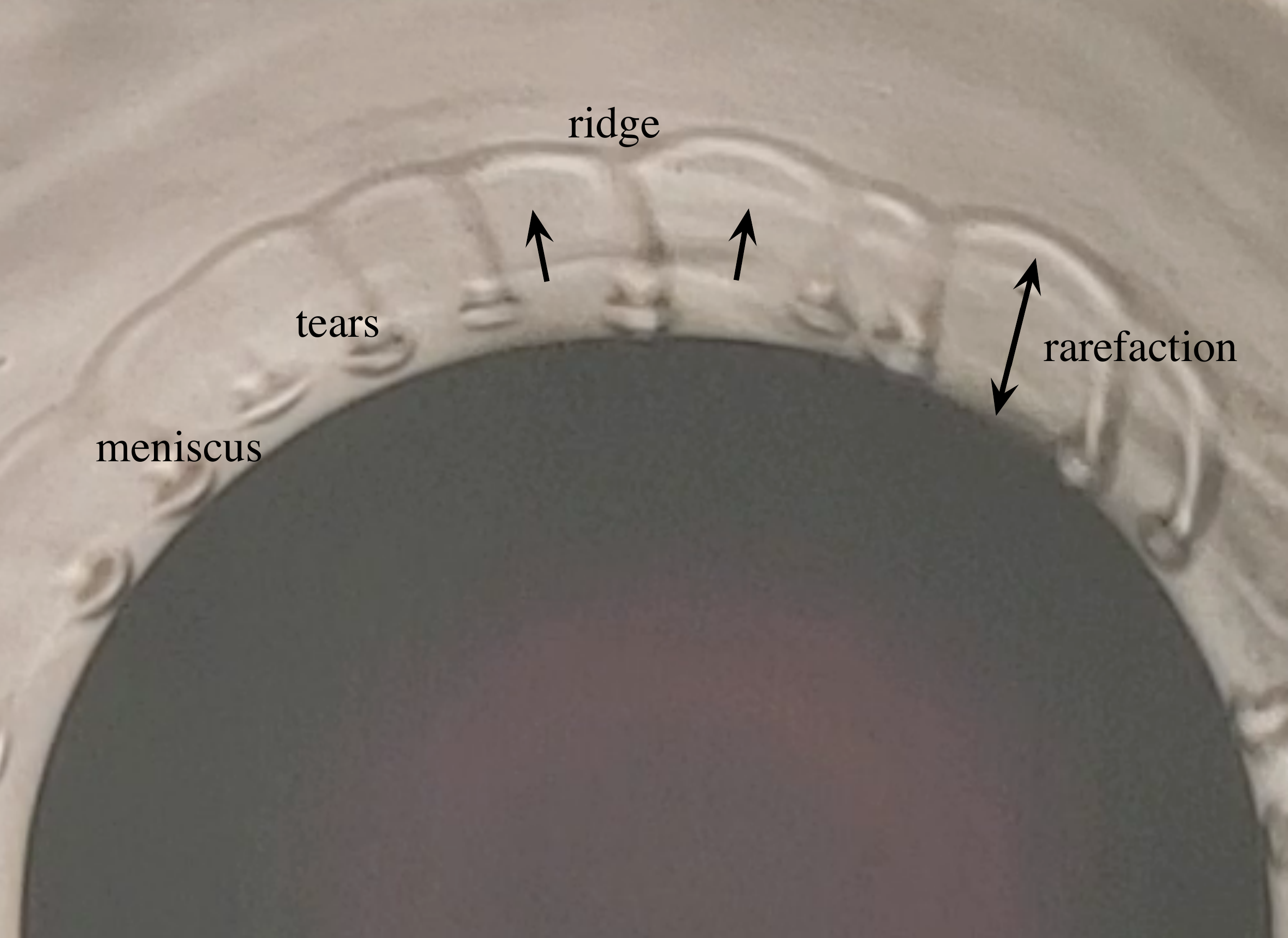}
\includegraphics[width=0.43\textwidth]{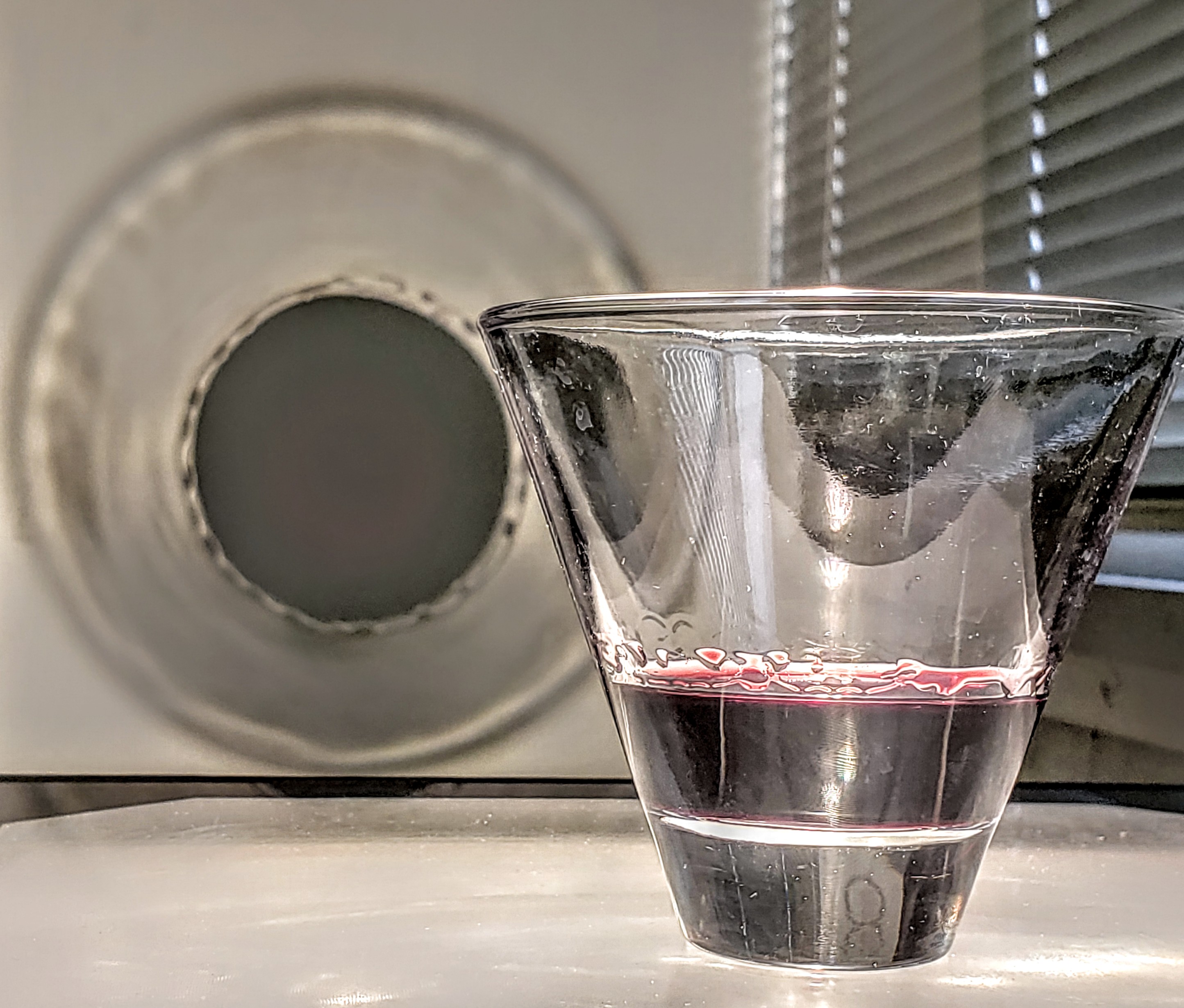}

\caption{Tears of wine experiment. (left) top view and (right) side view and projection of a stemless martini glass with inclination angle $\alpha = 65\degree$, using $18\%$ Port wine. Swirling the wine around the glass creates a front that forms out of the meniscus. The draining film advances up the glass and destabilizes into wine tears.} 
\label{fig:RUC_experiment_witharrows}
\end{figure}

It is difficult to reproduce many of the experiments performed with an initially dry surface discussed in the prior literature, because of the need to control the wetting properties of the contact line.  
Ordinary glassware will be affected by the way it is cleaned (e.g. see \cite{YouTube-AS2015} in which the author claims that glasses cleaned in a dishwasher with an additive to avoid spotting makes it more difficult to see wine tears).  
However, one can quickly observe wine tears by \red{actively pre-wetting (pre-swirling)} the glass as one would do when drinking a beverage or swirling the wine in the glass before drinking it.  
A \red{preswirled} glass can produce dramatic wine tears \red{(see \cite{YouTube-DQ2013} for an illustration)}. For the first time in the context of tears of wine, we identify the existence of another fundamental type of shock, the reverse undercompressive (RUC) shock, that involves a thicker film receding from a thinning region. Thin film structures involving an undercompressive leading shock and a trailing RUC shock were first identified in \cite{munch2003pinch,sur2003reverse} for dip-coating experiments with a thermal gradient that drives the film against gravity. The model used in \cite{munch2003pinch,sur2003reverse} is the same one we consider here.

Our experiments are performed using port wine of alcohol concentration $C = 18\%$ and a stemless martini glass of inclination angle $\alpha = 65\degree$ \red{in a room with controlled temperature at 75\degree F}. 
One can cover the glass immediately after pouring the wine, to temporarily suppress the evaporation of alcohol.
\red{A few seconds after pouring and covering, we give} the covered glass a brief \red{slow} swirl \red{ for about 3 seconds} and coat the substrate.  We observe that the initial swirl provides a 
surface with a thin draining film. We leave the cover on for \red{$\sim 10$} seconds until the swirl is no longer visible and the draining has settled down.  After removing the cover, evaporation quickly increases, inciting a ``reverse'' front to climb out of the meniscus,  followed by the formation of wine tears falling back into the bulk. \red{The experiments are highly reproducible and the times indicated here, as long as they are in the order of seconds, for swirling, covering, and uncovering, do not affect the outcome observed. This is supported by the theory for a range of preswirled thickness.} The forming front is characterized by a depression, i.e. the film ahead of the front is thicker than the film behind it.  It is in a sense, a ``dewetting'' front that leaves a thinner layer behind it.  The formation of the moving front is initiated by a pinch-off that occurs in the meniscus, as predicted in \cite{munch2003pinch}.
Snapshots of this experiment displayed in FIG.~\ref{fig:RUC_experiment} show a front that appears out of the meniscus and destabilizes into wine tears after $\sim 10$ seconds.
The left half of each image is a reflection, that is enhanced to visualize and capture the moving front. Around the center we observe a circular wave forming and travelling outward from the meniscus up the glass. The tears originate from the instability of that wave and drain back into the bulk fluid. Such waves appear to be the dominant behavior in the formation of the actual ``tears of wine". We note that the ``swirling'' initial condition may lead to different film thicknesses depending on the force of the swirl, i.e. the coating thickness is not quantitatively reproducible by manual swirling.  We now present a theory that shows that such film thicknesses, within a fairly broad range, all produce the same general pattern of a reverse undercompressive wave emerging from the meniscus, as in FIG.~\ref{fig:RUC_experiment_witharrows}. The predicted front behavior is universal within a range of coating thicknesses, to the point where one can do reproducible demonstrations at the dinner table.
Another signature that this is an RUC shock is that the tears emanate from the wave and travel downward, away from the shock, and towards the meniscus.  This is indicative of characteristics going through the wave, away from its direction of travel, because perturbations, to leading order, travel along characteristics.  A diagram for this type of behavior is shown in  FIG.~\ref{fig:characteristics} in the middle panel.  This is in contrast to a compressive wave in which disturbances, traveling along characteristics, enter the shock from both sides. 
One would expect instabilities of a compressive wave to travel with the wave, like in the case of the fingering instabilities seen in FIG.~\ref{fig:Ethanol_experiment}.

\begin{figure}
\centering
\includegraphics[width=7.7cm,height=5.2cm]{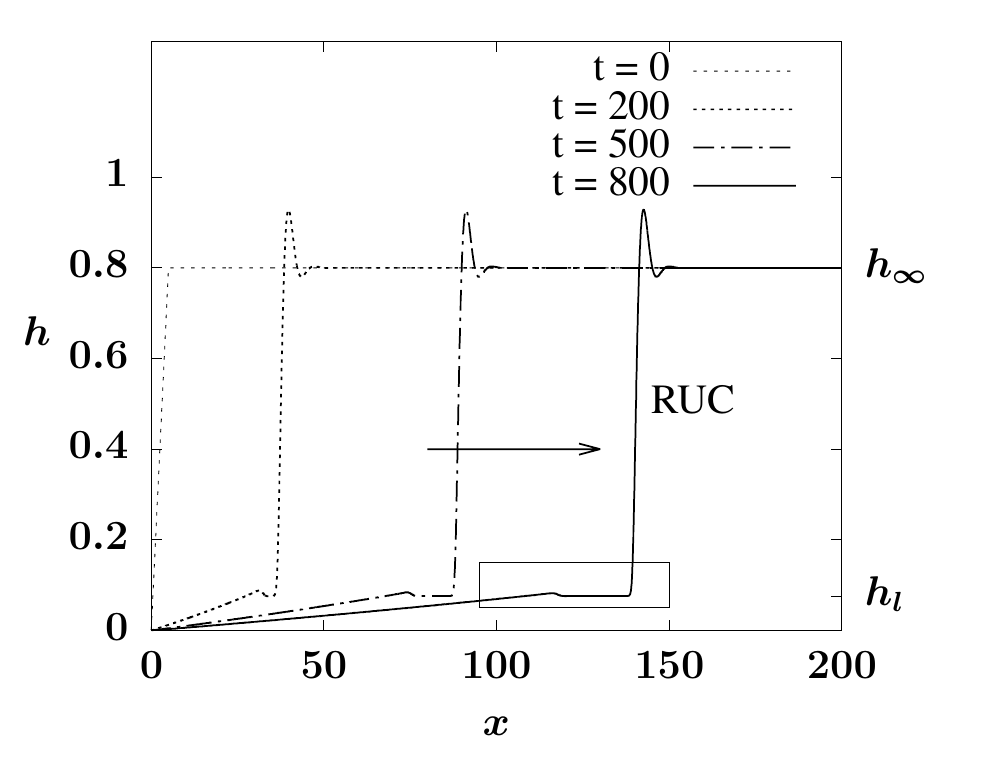}
\includegraphics[width=7.4cm, height=5.2cm]{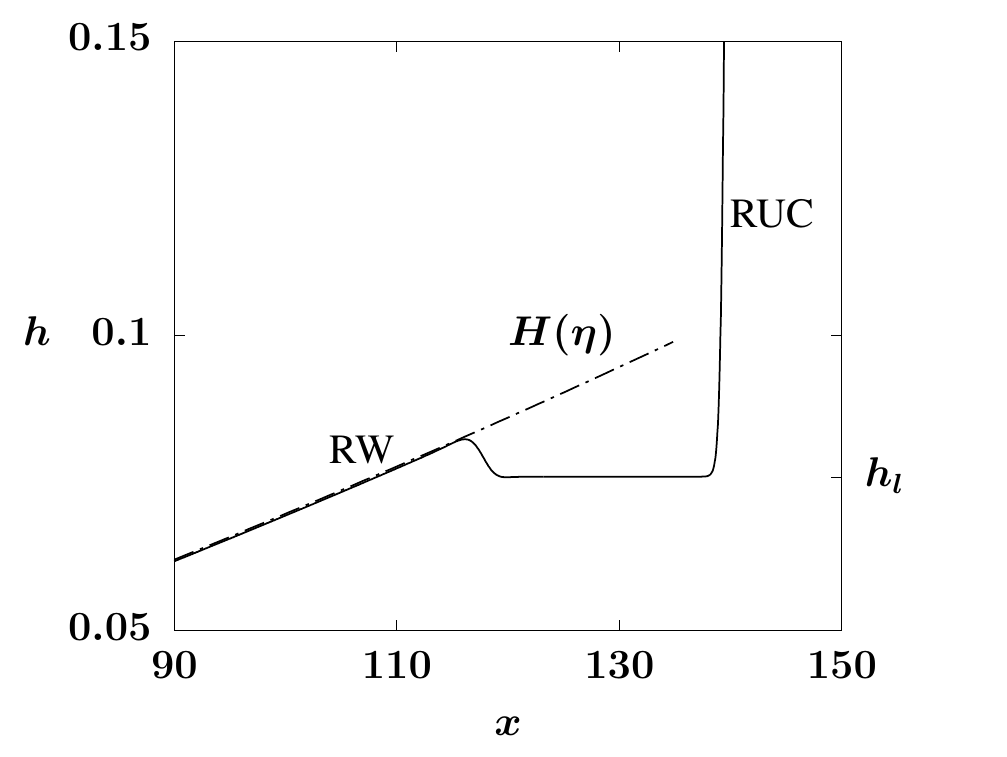}

\caption{(left) The formation of a reverse--undercompressive (RUC) shock with initial condition \eqref{eq:ic2} and boundary conditions \eqref{bc_swirl}. (right) A closeup view of the boxed region on  the left, at dimensionless time $t = 800$, showing a good agreement with the rarefaction wave (dot-dashed line) $H(\eta)$ in \eqref{rarefaction} for $\eta = x/t$.  The rarefaction separating from the RUC shock is a signature of a nonclassical shock, it is shown by the flat state $h_l$ in between the rarefaction and the RUC shock.
} 
\label{fig:RUC_piecewiseConstant}
\end{figure}

We now match the observed experimental behavior in FIG.~\ref{fig:RUC_experiment_witharrows} to solutions of the one dimensional model.
To approximate the initial profile of the draining film, after swirling, and immediately after the evaporation starts, we assume a pinch-off of the meniscus, as discussed in \cite{munch2003pinch}.  For simplicity we start with an initial condition that has steep constant slope jump connecting the meniscus to the coating layer,
\begin{equation}
    h_0(x) = 
    \begin{cases}
   h_{\infty} &\mbox{ for } x > x_L, \\
    h_{eq} + \frac{h_{\infty}-h_{eq}}{x_L} x &\mbox{ for } 0 < x \le x_L,
    \end{cases}
    \label{eq:ic2}
\end{equation}
where the film thickness $h_{eq}$ approximates the near-rupture film profile near the edge of the meniscus (as in FIG. 6 from \cite{vuilleumier1995tears}), and $h_{\infty}$ sets the thickness of the draining film due to the swirling of the glass.   We take $h_{\infty}$ to be independent of time however a more complete model could include a weak time dependence due to the dynamics further up the glass.
For the lower boundary conditions we apply
\begin{subequations}\label{bc_swirl}
\begin{equation}
    h(0) = h_{eq}, \qquad h_{xxx}(0) = 0,
\label{bc_meniscus}
\end{equation}
which assumes a fixed near-rupture film thickness and a zero curvature gradient at $x=0$. The upper boundary condition is
\begin{equation}
    h\to h_{\infty} \qquad \mbox{for } x \to \infty.
\label{bc_swirl_r}
\end{equation}
\end{subequations}
Our numerical simulations of equation \eqref{eqn:nondim} have $D = 0.0146$, $x_L = 5$, and $h_{eq} = 0.001$. These dimensionless parameters correspond to the dimensional values taken from \cite{venerus2015tears} (See in Appendix \ref{appendix:extended_survey} experiment (DI) of Table~\ref{tab:exp}) with a modified inclination angle $\alpha = 65\degree$. Unlike the cases discussed in Section \ref{sec:survey} where the film thickness $h_{\infty}$ is determined by the parameters $D$ and $b$ based on the meniscus dynamics, here we specify the value of $h_{\infty}$ to approximate the thickness of the initial draining film formed by the glass swirling in the experiment.
We will pick typical values for $h_{\infty}$ that correspond to a balance between the draining effect and the Marangoni stress.  We find that a wide range of such $h_{\infty}$ produce the same qualitative behavior.

\begin{figure}
\centering
\includegraphics[width=7.7cm,height=5.2cm]{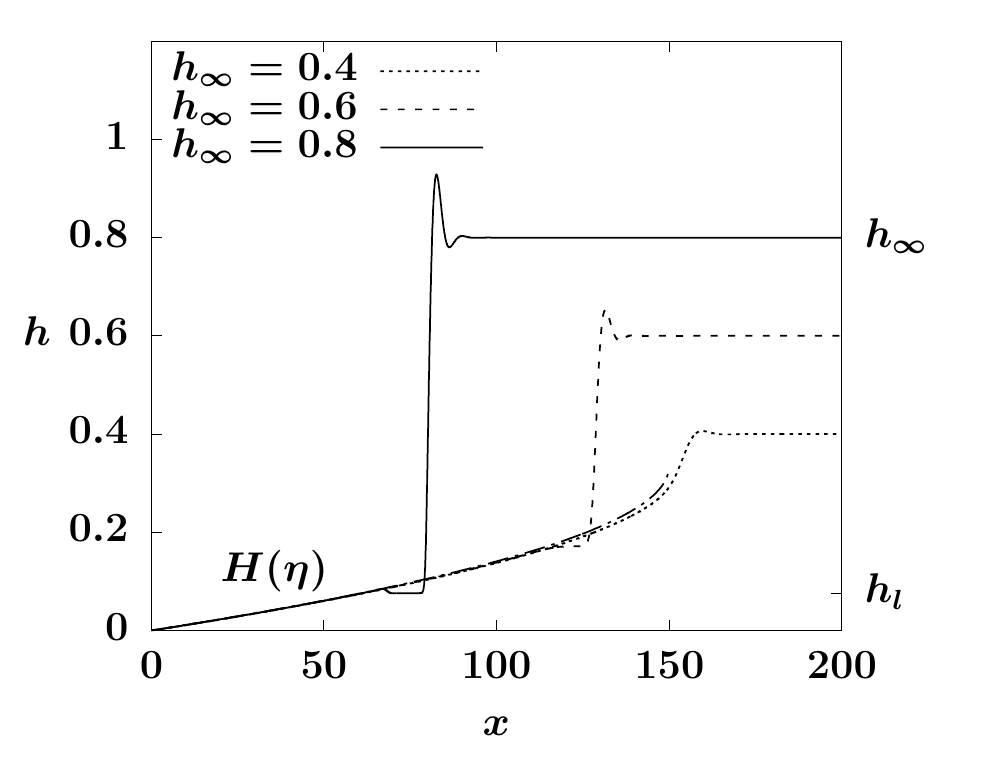}
\includegraphics[width=7.4cm, height=5.2cm]{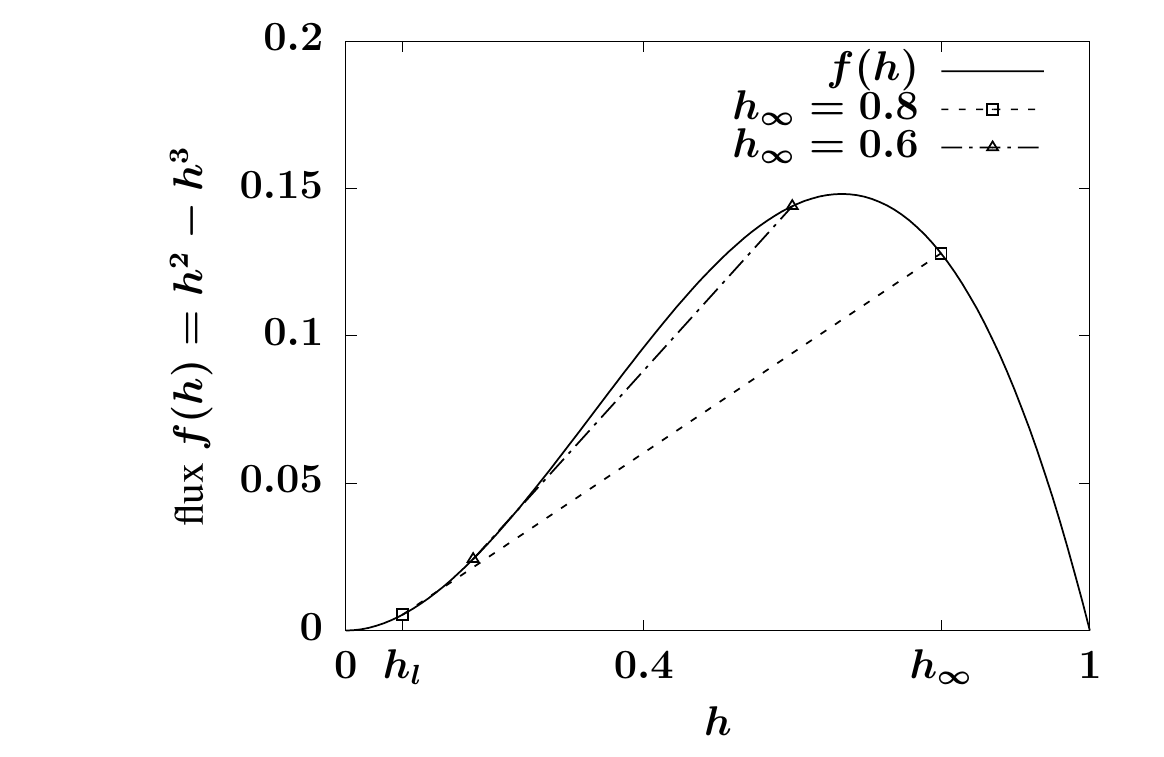}
\caption{(left) A comparison of shock solutions at $t = 450$ for initial data \eqref{eq:ic2} with varying $h_{\infty}$ showing reverse--undercompressive (RUC) shocks for $h_{\infty}=0.6, 0.8$, and a single rarefaction wave for $h_{\infty} = 0.4$.
The critical thicknesses $h_{\infty}$ and $h_l$ are marked for the $h_{\infty} = 0.8$ profile.  Note that the rarefaction part of the solution is independent of $h_{\infty}$.
(right) The flux diagram with two undercompressive connections for the RUC shocks with $h_{\infty} = 0.6, 0.8$.} 
\label{fig:RUC}
\end{figure}

FIG.~\ref{fig:RUC_piecewiseConstant} (left) shows a typical numerical simulation of the model \eqref{eqn:nondim} for the evolution of the film height starting from the initial condition \eqref{eq:ic2} with $h_{\infty} = 0.8$.
The left-hand boundary models the pinchoff at the meniscus, a phenomenon that has been widely studied in coating films \cite{munch2003pinch,carles1993thickness}, and is driven by the dynamics of the meniscus as it forms the equilibrium height $h_{eq}$. This pinchoff leads to a pronounced complex wave form emanating from the meniscus.  
In the early stage of the dynamics, we see a double wave structure emerging. There is a rarefaction fan near the meniscus and a shock wave connecting to the larger height $h_{\infty}$.  Note that the two waves separate from each other as time passes. A flat film of thickness $h=h_{l}$ (see the tick labels on the right vertical axes in FIG.~\ref{fig:RUC_piecewiseConstant})
connects the right edge of the rarefaction wave and the left edge of the leading wave.  
This is typical for such double wave structures involving undercompressive waves --- a new equilibrium height emerges that is driven by the solution on each side \cite{munch2003pinch,bertozzi1999undercompressive}.
 FIG.~\ref{fig:RUC_piecewiseConstant} (right) shows a close-up of the solution profile at $t = 800$ delimited by a box in FIG.~\ref{fig:RUC_piecewiseConstant} (left), indicating that the rarefaction wave (RW) portion of the solution is given by $h(x,t) = H(x/t)$ in \eqref{rarefaction}.   To further verify the UC structure, we plot the connection between $h_l$ and $h_{\infty}$ on the flux function diagram (see FIG.~\ref{fig:RUC}, right panel).  The chord crosses the graph of the flux function, illustrating that the shock violates the entropy condition.

\begin{figure}
\centering
\includegraphics[width=7.7cm,height=5.2cm]{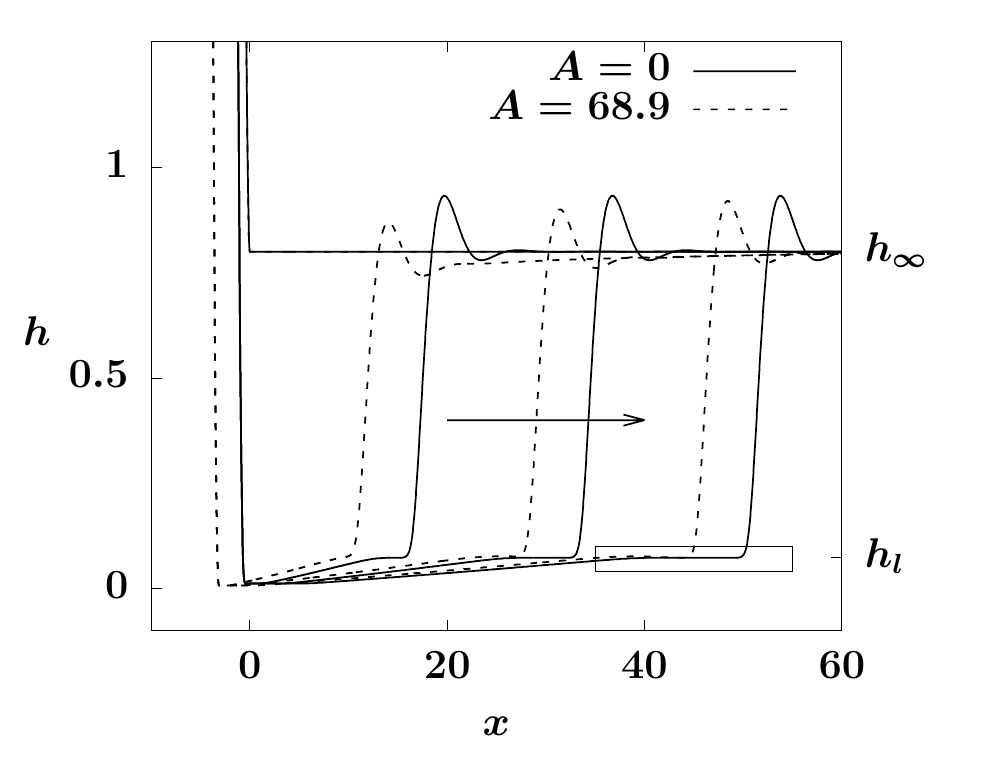}
\includegraphics[width=7.4cm, height=5.2cm]{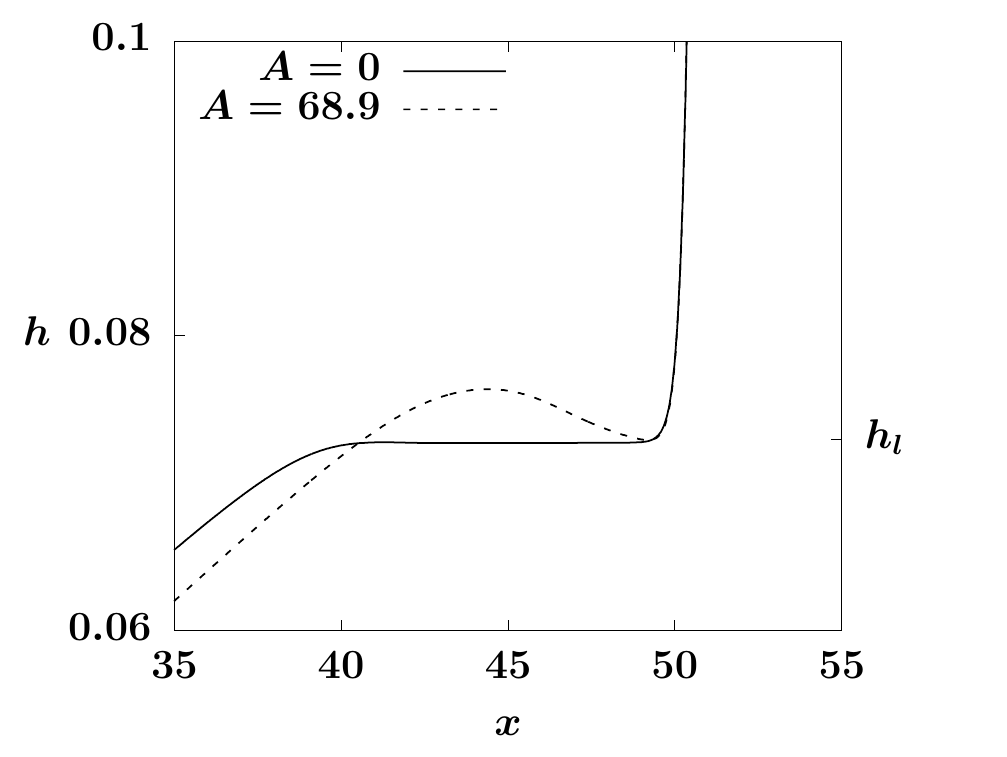}
\caption{
(left) Advancing waves starting from initial condition \eqref{ic} at times $t = 0, 100, 200, 300$ governed by \eqref{modelcurvature} with meniscus boundary conditions \eqref{eq:meniscus_bc}, showing a comparison of fronts influenced by curvature effects $(A=68.9, x_0 = 25)$ and without curvature effects $(A = 0)$. (right) A closeup view of the boxed region on the left at time time $t = 300$ with the curve for $A = 68.9$ shifted by $\Delta x = 5.3$. The other settings are identical to those in FIG.~\ref{fig:RUC_piecewiseConstant}.} 
\label{fig:RUC_parameter_comparison}
\end{figure}

    FIG.~\ref{fig:RUC} shows film heights with  $h_{\infty}= 0.4, 0.6,0.8$. 
    For $h_{\infty}=0.8$, we have $h_l \approx 0.076$; for $h_{\infty}=0.6$, $h_l \approx 0.17$.
In both cases the shock violates the entropy condition.  For $h_{\infty}= 0.4$ the dynamics is dominated by the rarefaction fan which terminates abruptly in the flat film on the right hand side without the pronounced capillary ridge seen in the other two cases.  In this case, the dynamics are dominated by the Marangoni stress. For really thick draining films (e.g. with $h_{\infty}>1$), the dynamics is dominated by gravity so we expect a range of $h_{\infty}$ for which this phenomenon occurs. 
Incorporating the conical-shaped substrate curvature effects and the meniscus dynamics also influences the profile of the RUC shock. Here we combine these effects by using model \eqref{modelcurvature} with the meniscus boundary conditions \eqref{eq:meniscus_bc}. Starting from the initial condition \eqref{ic} that emulates the meniscus profile, in FIG.~\ref{fig:RUC_parameter_comparison} (left) we plot the simulation results with $A = 68.9$, $x_0 = 25$ against the profiles without curvature effects ($A=0$). Other system parameters are set to be $(D, h_{\infty}) = (0.0146, 0.8)$ which match the simulation shown in FIG.~\ref{fig:RUC_piecewiseConstant}. This comparison shows that the early stage pinch-off near the meniscus is sensitive to the substrate curvature effects, which leads to a different stable meniscus profile and location where the near-rupture film thickness $h_{eq}$ is attained. This difference leads to a spatial shift in the later stage dynamics, whereas the rarefaction wave and the speed of the moving front do not change significantly. A closeup view of the wave fronts at $t = 300$ is also shown in FIG.~\ref{fig:RUC_parameter_comparison} (right), where the curve for $A = 68.9$ is horizontally shifted to align with the $A = 0$ curve. It indicates that the RUC shock obtained for $A=0$ is less pronounced with the presence of weak substrate curvature effects.

Previously, it has been shown that the RUC wave is unstable with respect to transverse perturbations \cite{munch2003pinch}. As the wave destabilizes, the transverse perturbations enter the space between the RUC wave and the rarefaction fan, which agrees with the wine tears being shed downward from the rising \red{circular} wave in our experiment (see FIG.~\ref{fig:RUC_experiment_witharrows}). As time goes on, the tears travel into the rarefaction fan and get elongated as the rarefaction wave expands.
The theory here suggests a mechanism for the onset of the wine tears. A fully nonlinear 2D simulation of the model could be done in future work to understand the longer time dynamics of the wine tears.

\section{Conclusion}
\label{sec:conclusion}
In this paper we introduce a model for  the tears of wine phenomena that describes the balance between gravity and a Marangoni stress induced from alcohol evaporation. 
 The dynamic model is the same equation that has been used to describe thermally driven films balanced by gravity.  This work is the first to connect that literature to the tears of wine problem.  We argue that the actual wine tears, which drain down the glass, in contrast to the well-known fingering instability of driven fronts, which travel in the same direction of the front, arise from an instability of a reverse undercompressive shock.  They can be easily observed by prewetting the glass as one would do in the context of drinking a beverage or swirling the wine around the glass.  We are able to create fairly reproducible experiments of this phenomenon by pre-swirling the glass, while covered, to suppress evaporation.  Removing the cover, after the initial pre-swirl, leads to a circular wave emanating from the meniscus that quickly destabilizes into downward draining wine tears.  
 
Our main model is for a flat substrate.  This model allows for easy identification of different wave forms because they have an exact self-similar structure.  
We also show that incorporating the substrate curvature effects into the governing equation can lead to dynamic behaviors that are qualitatively similar, and the difference can be quantified through numerical simulations.

It has been shown in the literature \cite{bertozzi1999undercompressive, bertozzi1998contact} that while the undercompressive shocks are stable, the compressive shocks and reverse undercompressive shocks are unstable to fingering \cite{munch2003pinch}. More work could be done to quantitatively predict the spacing of the wine tears observed in these experiments. This would involve analyzing the linear stability of the RUC ridge along with fully nonlinear 2D numerical simulations.

Prior experimental results presented in Section \ref{sec:survey} illustrate the formation of different shock structures under different experimental conditions. For example we observe that the surface tension gradient $\tau$ and the precursor height $b$ are pivotal to the formations of different shocks. 
While a conical-shaped martini glass is easy to model because of its constant inclination angle, one could also incorporate three-dimensional complex surface geometry to the model such as that observed in common wine glasses. We believe a more accurate description of the phenomenon may be obtained via a careful consideration of the three-dimensional geometry and the surface tension gradient.
Finally, we note that our model \eqref{eqn:nondim} assumes a constant surface tension gradient.
As the film climbs up, this assumption eventually fails, requiring a modification of the model to describe the full dynamics.
Along these lines, downward draining wine tears can also be observed from fluid that accumulates at the top of the glass, forming a stationary capillary ridge \cite{YouTube-C2015,nikolov2018tears,venerus2015tears}. It would be interesting to try to model the formation of this structure.

\subsection*{Acknowledgments}
We would like to extend special thanks to Andreas M{\"u}nch for supporting the idea and helping implement the meniscus model as well as for many thoughtful comments.
We would like to also extend our thanks to Anne Marie Cazabat for helpful discussions when conducting this research.
This material is based upon work supported by Simons Foundation Math+X investigator award number 510776 and YD was supported by the National Science Foundation Graduate Research Fellowship under Grant No. DGE-1650604.

\bibliography{tow}
\raggedbottom

\pagebreak
\appendix

\section{Extended Survey of Prior Experimental Works}
\label{appendix:extended_survey}
Here we review the existing experimental literature summarized in TABLE \ref{tab:exp} and \ref{tab:exp2}.
The works discussed are:
(A) ``Tears of Wine'' \cite{fournier1992tears}, with alcohol concentration $C = 70\%$,
(B) ``Tears of wine: the stationary state''\cite{vuilleumier1995tears}  with experiments (BI) of data taken from Figure 5b of their paper with a curvature-driven film and $C = 70 \%$. and (BII) of data taken from Figure 5b, now with a gravity-driven regime and $C = 70\%$. As mentioned in section \ref{sec:survey}, (BI) and (BII) refer to the same physical setting with different assumptions when calculating the surface tension gradient $\tau$. (C) ``Evaporative Instabilities in Thin Films'' \cite{hosoi2001evaporative} experiment (CI) refers to the settings described in Table 3 and Figure 10 of \cite{hosoi2001evaporative}. Experiment (CII) refers to Table 3 in \cite{hosoi2001evaporative} but in addition uses the experimental settings given in Figure 11 of the same paper. (D) ``Tears of wine: new theory on an old phenomena'' \cite{venerus2015tears} presents two experiments, for wine and cognac. We denote the experiments as (DI) and (DII). 

In the tables, $\bdim$ refers to the precursor thin film height measured in $\mu m$, $\hdim$ $(\mu m)$ refers to the height of the film at the bulk (of the thin film), $\gamma$ refers to the surface tension ($N/m$), $\tau$ ($Pa$) refers to the surface tension gradient, $\alpha$ is the inclination angle measured in degrees, $\mu$ is the dynamic viscosity of the film (mili- $Pa~ s$), and $C$ is the volumetric water-ethanol fraction. The collection of symbols and typical dimensional values are also presented in TABLE \ref{tab:symbols} for convenience.

In the third column of the tables of the experiments we present dimensionless values for $(D, b)$ that appear in the PDE model in equation \eqref{eqn:nondim}. We remark that some of the values we present are interpolated from other experiments.
For example, in Table \ref{tab:exp}, the dimensional precursor value $\bdim$ is only provided in the literature for experiment (A). 
For simplicity, we use the dimensional precursor thickness $\bdim = 2\mu$m of (A) for the other experiments with high alcohol concentrations and low inclination angles  (experiments, (BI), (BII), and (CI)).
For experiments with higher inclination angles and lower alcohol concentrations (experiments (CII), (DI), and (DII)), the authors did not report the precursor height and we may not interpolate it since we do not have $\bdim$ measurements for such settings.

\begin{table*}

    \centering
        \hrule

    \begin{tabular}{ c c c c }
       \textbf{Physical Quantity} & ~~~\textbf{Symbol} ~~~ & ~~~ \textbf{Typical dim. value} ~~~ & \textbf{Dimensionless range}   \\
         Upstream thickness & $\hdim, h_{\infty}$ & $30 \mu m \textrm{--} 98 \mu m$ & $0 \textrm{--} 2.1$ \\
    Precursor thickness &  $\bdim, b$ & $2\mu m$ & $10^{-2} \textrm{--} 10^{-1}$ \\
    Surface tension & $\gamma$ & $22.39mN /m$ \textrm{--} $72.86 mN /m$   \\
    Surface tension gradient & $\tau $ \\ 
    Inclination angle &  $\alpha$ & $7^{\circ} \textrm{--} 45^{\circ} $\\ 
    Fluid dynamic viscosity & $\mu$ & $ 1.1 mPa~s^{\S}$ \\ 
    Alcohol concentration & $C$  & $0.15 \textrm{--}0.7$\\
    Density & $\rho$ & $784 kg /m^3 - 973 kg /m^3$ \\ 
    \end{tabular}

    \caption{Relevant dimensional groups used in TABLE \ref{tab:exp}}
                   \hrule

    \label{tab:symbols}
\end{table*}

\begin{table*}
\caption{Experimental results from literature and corresponding theory }

\resizebox{1\textwidth}{!}{
\begin{tabular}{ | c | c | c | c |}

\hline
\makecell{\large\textbf{Experiment}} & \makecell{\large \textbf{ Dimensional constants}
} & \makecell{\large\textbf{Dimensionless} \large\textbf{constants} 
} & \large\textbf{Images of experimental results}\\
\hline
\makecell{(A) Tears of Wine \\
(Fournier and Cazabat)
\cite{fournier1992tears}} & \makecell{$\hdim = 55 \mu m $ \\
$ ~~ \bdim = 2 \mu m $\\ $\gamma = 0.0298 N/m^{**}$ \\
$\tau = 0.055 Pa $\\
$\alpha = 9^{\circ}$ \\
$\mu = 2.1 m Pa ~ s ^{\S}$ \\ $C = 0.7$ \\ $\rho = 852 ~kg / m^3$} & \makecell{
$b = 0.0317 $\\ 
$D  = 0.353 $
}
&\makecell{\includegraphics[width = 8cm]{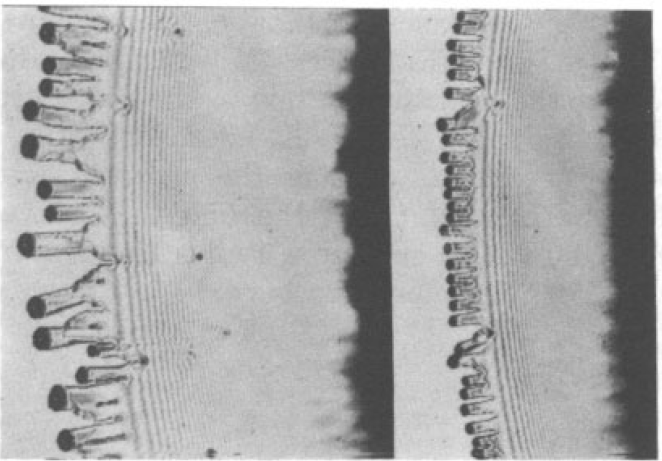}\\  Experiment (A) settings at different times}\\ 

\hline
\makecell { (CI) Evaporative \\ instabilities \\ in climbing films\\
experiment I \\ (Hosoi and Bush)
\cite{hosoi2001evaporative} } & \makecell{$\hdim = 30 \mu m $ \\
$ ~~ \bdim  = 2 \mu m \dagger$\\ 
$\gamma = 0.027 N/m$ \\
$\tau = 0.025 Pa $\\
$\alpha = 4\degree$\\
$\mu = 1 m Pa ~ s^{\S}$ \\
$C = 0.65-0.7$ \\ $\rho = 852 ~kg / m^3$ } 
& \makecell{
$b = 0.031 $\\ 
$D  = 0.639$
} & \makecell{\includegraphics[width = 8cm]{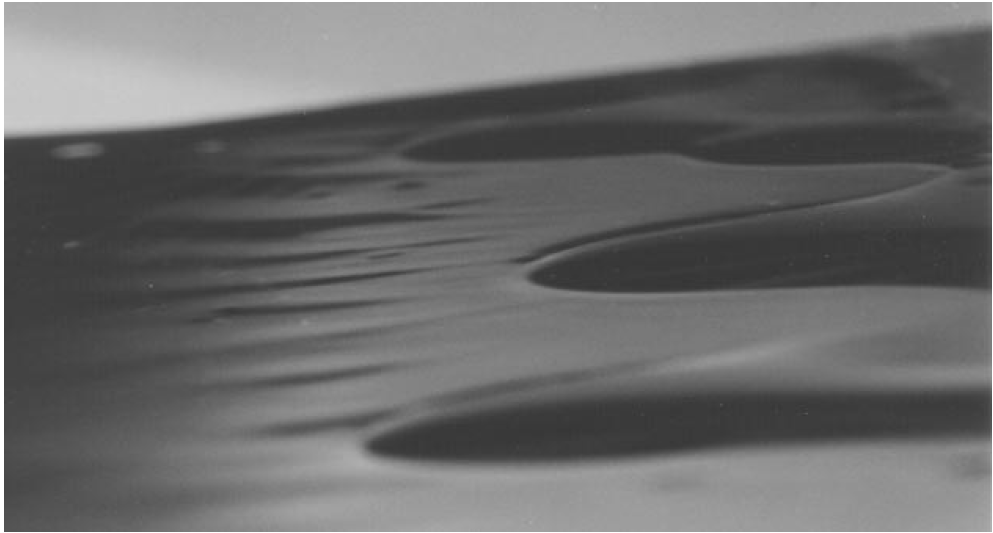}\\ $\alpha = 4^{\degree} $ Concentration as in (CI) Reprinted with permission from \cite{hosoi2001evaporative} \\ and the Cambridge University Press.}\\

\hline
\makecell{(DI) Tears of wine: \\
new insights on an\\
old phenomenon\\
(wine) \\(Venerus and Simavilla)
\cite{venerus2015tears} } 
& \makecell{
$\gamma = 0.054N/m^{**}$ \\
$\tau = 0.08 Pa $\\
$\alpha = 45 ^{\circ}$ \\
$\mu = 1.1m Pa ~ s^{\S}$ \\ $C = 0.13$ \\ $\rho = 973 ~kg / m^3$} &

\makecell{
$D  = 0.0338 $ \\
\\} 
&  \makecell{\includegraphics[width = 8cm]{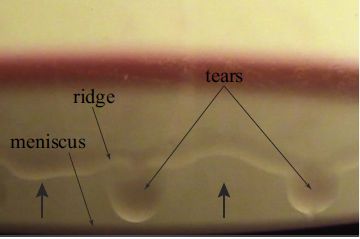} \\ Image of experiment (DI) $\mathparagraph$}\\ 
\hline
\makecell{Our experiment \\ shown in FIG.~\ref{fig:Ethanol_experiment}} 
& \makecell{
$ ~~ \bdim = 2 \mu m  \dagger$\\ 
$\gamma = 0.0298N/m^{**}$ \\
$\tau = 0.055 Pa $\\
$\alpha = 9 ^{\circ}$ \\
$\mu = 2.1m Pa~ s^{\S}$ \\ $C = 0.7$ \\ $\rho = 852 ~kg / m^3$\\
Set-up matches (A) \\
} &
\makecell{ $b =  0.0317$ \\ 
$D  = 0.353 $ 
} &  \makecell{\includegraphics[width = 6cm]{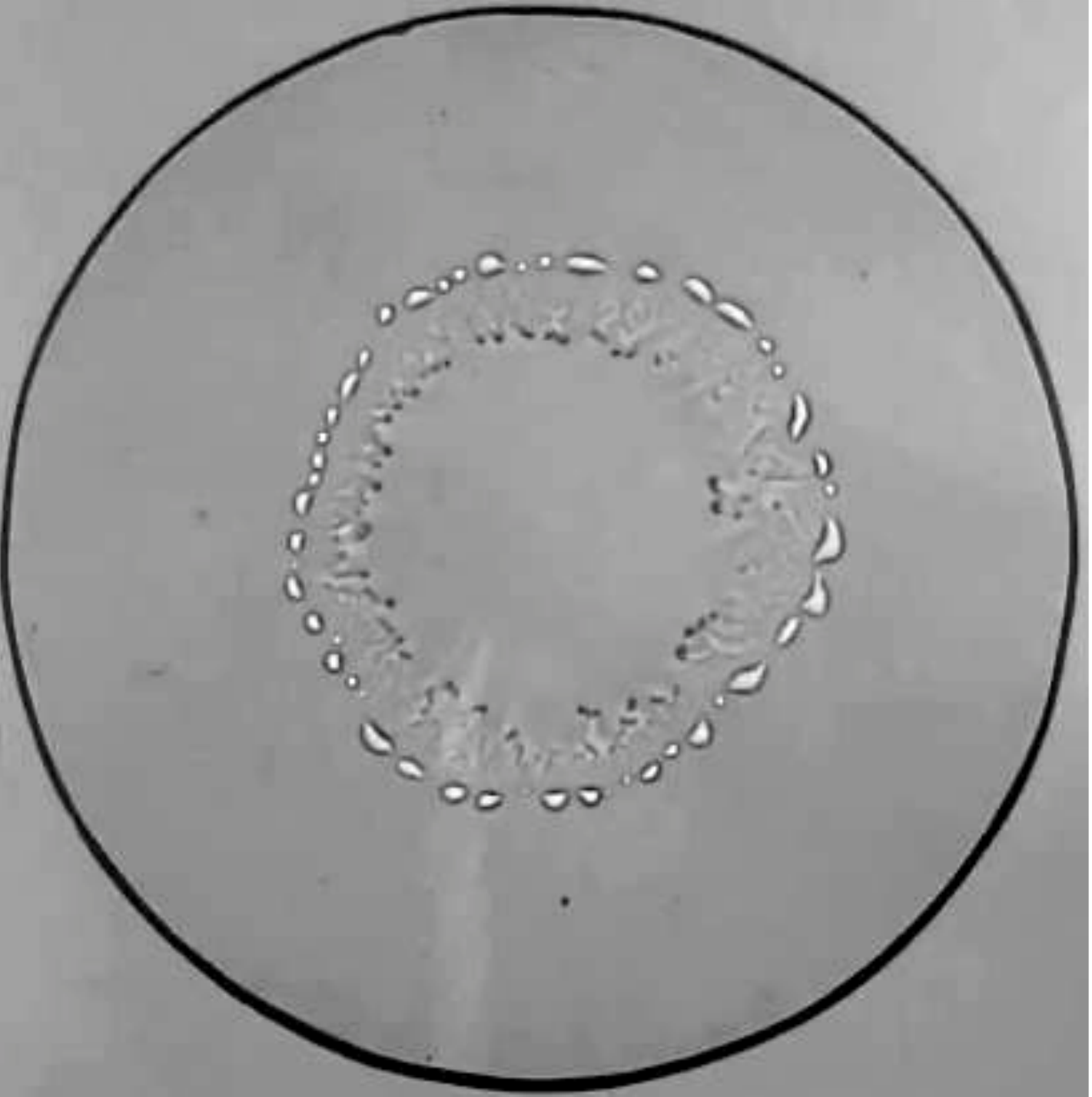}} \\ 

\hline
\end{tabular}
}

{\scriptsize $^{**}$ refers to surface tension interpolated from \cite{vazquez1995surface}\qquad
$^{\S}$ refers to viscosity interpolated from \cite{Visc}\\
$\dagger$ refers to precursor thickness taken from \cite{fournier1992tears} \qquad
$\mathparagraph$ Reproduced from
``Tears of wine: new insights on an old phenomenon'' 
\cite{venerus2015tears} under compliance with the creative commons 4.0 licence.  }
  \label{tab:exp}
\end{table*}


\begin{center}
\begin{table}
\caption{Additional experimental results from literature and corresponding theory}
\resizebox{1\linewidth}{!}{
\begin{tabular}{ | c | c | c |}
\hline
\textbf{Experiments} & \makecell{ \textbf{ Dimensional constants}
} & \makecell{\textbf{Dimensionless constants} 
} \\ 
\hline

\makecell{(CII) Evaporative instabilities in \\
climbing films
experiment II \\ (Hosoi and Bush)
\cite{hosoi2001evaporative} } & \makecell{
$\gamma = 0.027 N/m$ \\
$\tau = 0.025 Pa $,~
$\alpha = 20.05 ^{\circ}$ \\
$\mu = 1 m Pa~ s^{\S}$ ,~ $C = 0.65-0.7$ \\ $\rho = 852 ~kg / m^3$} & \makecell{
$D  = 0.072 $}\\

\hline
\makecell{(BI) Tear of wine: \\ The stationary state \\
experiment I \\ (Vuilleumier et al.)
\cite{vuilleumier1995tears} } & \makecell{$\hdim = 98 \mu m $,
$ ~~ \bdim  = 2 \mu m \dagger$\\ $\gamma = 0.0298 N/m^{**}
 ~,~\tau = 0.033 Pa $\\
$\alpha = 6 ^{\circ}~,~\mu = 2.1 m Pa ~ s^{\S}$ \\ $C = 0.7~,~\rho = 852 ~kg / m^3$} &
\makecell{
 $b =  0.0353$\\ 
$D  = 0.43 $
}  \\

\hline
\makecell{(BII) Tear of wine: \\The stationary state \\
experiment II \\ (Vuilleumier et al.)
\cite{vuilleumier1995tears} } & \makecell{
    $ \bdim  = 2 \mu m \dagger$, ~ $\gamma =0.0298 N/m^{**}$ \\
$\tau = 0.10 Pa $,~
$\alpha = 6 ^{\circ}$ \\
$\mu = 2.1 m Pa ~ s^{\S}$, ~ $C = 0.7$ \\ $\rho = 852 ~kg / m^3$} &

\makecell{
 $b = 0.0106 $\\ 
$D  = 0.966 $ 
}  \\ 

\hline
\makecell{(DII) Tears of wine: 
new insights \\on an
old phenomenon\\
(cognac) \\(Venerus and Simavilla)
\cite{venerus2015tears} } & \makecell{
$\gamma = 0.032N/m^{**}$ \\
$\tau = 0.06 Pa $,~
$\alpha = 45 ^{\circ}$ \\
$\mu = 2.35m Pa~s^{\S}$, ~$C = 0.35$ \\ $\rho = 926 ~kg / m^3$} &

\makecell{
$D  = 0.0346 $ \\
}  \\

\hline
\end{tabular}
\label{tab:exp2}
 }

{\scriptsize $^{**}$ refers to surface tension interpolated from \cite{vazquez1995surface}\qquad
$\S$ refers to viscosity interpolated from \cite{Visc}\\
$\dagger$ refers to precursor height taken from \cite{fournier1992tears}\\
}
\end{table}
\end{center}
\end{document}